\documentclass{article}
\usepackage[letterpaper,top=1in,bottom=1in,left=1in,right=1in]{geometry}
\usepackage[utf8]{inputenc}
\pdfoutput=1
\usepackage[authoryear]{natbib}
\usepackage{hyperref}
\hypersetup{colorlinks, linkcolor = black, citecolor = blue, urlcolor = blue, hypertexnames}
\usepackage{amsmath}
\usepackage{graphicx}
\usepackage{booktabs, multirow}
\usepackage{pdflscape}
\usepackage{caption}

\title{Low Frequency Marsquakes and Where to Find Them: Back Azimuth Determination Using a Polarization Analysis Approach}
\author{G\'{e}raldine Zenh\"ausern$^{1*}$, Simon C. St\"ahler$^1$, John F. Clinton$^2$, Domenico Giardini$^1$,\\ Savas Ceylan$^1$, Rapha\"{e}l F. Garcia$^3$}
\date{}

\begin{document}

\maketitle
\noindent
$^{*}$ Corresponding author: G\'{e}raldine Zenh\"ausern,
Institute of Geophysics,
ETH Zurich,
NO E 17,
Sonneggstrasse 5,
8092 Zurich,
Switzerland,
\url{geraldine.zenhaeusern@erdw.ethz.ch}\\
$^{1}$ Institute of Geophysics, ETH Zurich, Zurich, Switzerland \\
$^{2}$ Swiss Seismological Service, ETH Zurich, Zurich, Switzerland \\
$^{3}$ Institut Sup\'erieur de l'A\'eronautique et de l'Espace SUPAERO, Toulouse, France  \\

\section*{Abstract}
NASA's InSight mission on Mars continues to record seismic data over 3 years after landing, and to date, over a thousand marsquakes have been identified. With only a single seismic station, the determination of the epicentral location is far more challenging than on Earth. The Marsquake Service (MQS) produces seismicity catalogues from data collected by InSight and provides distance and back azimuth estimates when these can be reliably determined - when both are available these are combined to provide a location. Currently, MQS do not assign a back azimuth to the vast majority of marsquakes. In this work we develop and apply a polarization analysis method to determine the back azimuth of seismic events from the polarization of observed P and S-wave arrivals. The method is first applied to synthetic marsquakes, and then calibrated using a set of well-located earthquakes that have been recorded in Tennant Creek, Australia. We find that the back azimuth is estimated reliably using our polarization method. The same approach is then used for a set of high quality marsquakes recorded up to October 2021. We are able to estimate back azimuths for 24 marsquakes, 16 of these without MQS back azimuths. We locate most events to the east of InSight, in the general region of  Cerberus Fossae.

\section{Introduction}
NASA's Interior Exploration using Seismic Investigations, Geodesy and Heat Transport (InSight) mission landed on Mars in November 2018 and successfully completed the deployment of the Seismic Experiment for Interior Structure (SEIS) in February 2019 \citep{banerdt_initial_2020}. It consists of two sensors, the Very Broadband (VBB) and Short Period (SP) seismometers \citep{lognonne2019seis}. The lander is located in western Elysium Planitia. Figure \ref{Fig:Map} shows a global map of Mars, with a zoom inset of a region close to the lander. North of InSight lies the volcanic region Elysium Mons, and in north-eastern to eastern direction is the extensive and young Cerberus Fossae graben system \citep{berman_recent_2002}. The dichotomy boundary, separating the highlands of the southern hemisphere from the lowlands of the northern hemisphere, is near the landing site to the south. For an overview of the geology around the InSight landing site, see \citet{Golombek2020geology}. 

Since deployment, the VBB has recorded hundreds of seismic events \citep{giardini_seismicity_2020,clinton2021_catalog}. While the Apollo mission had up to four simultaneously recording stations on the Moon \citep{lammlein_lunar_1977}, there is only a single station on Mars. This makes event detection and location particularly challenging and means that classical seismological triangulation methods cannot be applied. \citet{bose_probabilistic_2017} propose a probabilistic single-station methodology to locate marsquakes that estimates distance and back azimuth separately, both from body wave and surface wave arrival times and motion direction. On Earth, back azimuths are not routinely determined by polarization, but similar methods are used to confirm sensor orientation, specifically for ocean bottom seismometers, whose orientation is otherwise unknown \citep[e.g. ][]{anderson_ocean_1987, stachnik_determination_2012}. Such methods typically use either the P- or the Rayleigh wave motion, since both are expected to occur in the plane pointing towards an earthquake. Deviations of both P-wave \citep{fontaine_upper_2009} and Rayleigh wave azimuth \citep{laske_global_1995} from great circle direction due to three-dimensional structure are observed, but are typically below 5$^{\circ}$. Having been deployed on the surface, the orientation of the SEIS instrument was determined using a sundial on its cover \citep{savoie_finding_2020} with an accuracy of 2.5$^{\circ}$. The instrument coordinate system is therefore known well enough to use polarization for marsquake location.

As described in \citet{clinton2021_catalog}, since no surface waves have been identified on Mars and the waveforms are affected by strong scattering \citep{lognonne_constraints_2020}, Rayleigh wave-based methods cannot be applied to Mars. For 8 marsquakes, the P-wave direction of motion could be used to determine the back azimuth, but the majority of marsquakes only have distances (obtained from S-P times), and subsequently, lack a location. Any analysis of seismicity and tectonics on Mars would greatly benefit if more marsquake locations were available. Similarly, moment tensors require a back azimuth to provide results \citep{Brinkman2021}. In \citet{martian_core_2021}, SM 4, a method based on the correlation and instantaneous phase coherence between vertical and horizontal components as a function of azimuth was used to locate a number of additional marsquakes.

In this paper we present an alternative way to estimate the back azimuth for marsquakes, which is based on the polarization of the signal obtained from eigenanalysis of the spectral matrix \citep{schimmel_polarized_2011, Samson1980_polarization}, allowing the identification of polarized signals even in presence of strong, complex noise. A similar method was previously used for orientation of ocean-bottom seismometers focusing on Rayleigh waves \citep{scholz_orienting_2017}. We show that it can be used jointly on P- and S-waves, yielding a robust and consistent back azimuth estimate.

We briefly summarize the existing approach used by the Marsquake Service \citep[MQS, ][]{Clinton2018MQS} to determine the back azimuth, and then describe the specifics of our proposed method. The method is then tested on synthetic martian data as well as a number of earthquakes. We finally apply this method to a set of high quality events recorded by InSight until end of September 2021 \citep[Marsquake Catalog Version 9, ][]{MQSCatalog} to determine their back azimuth and create a new and significantly extended set of marsquake locations.

\begin{landscape}
\centering
\begin{figure}[ht]
\noindent\includegraphics[width=24cm]{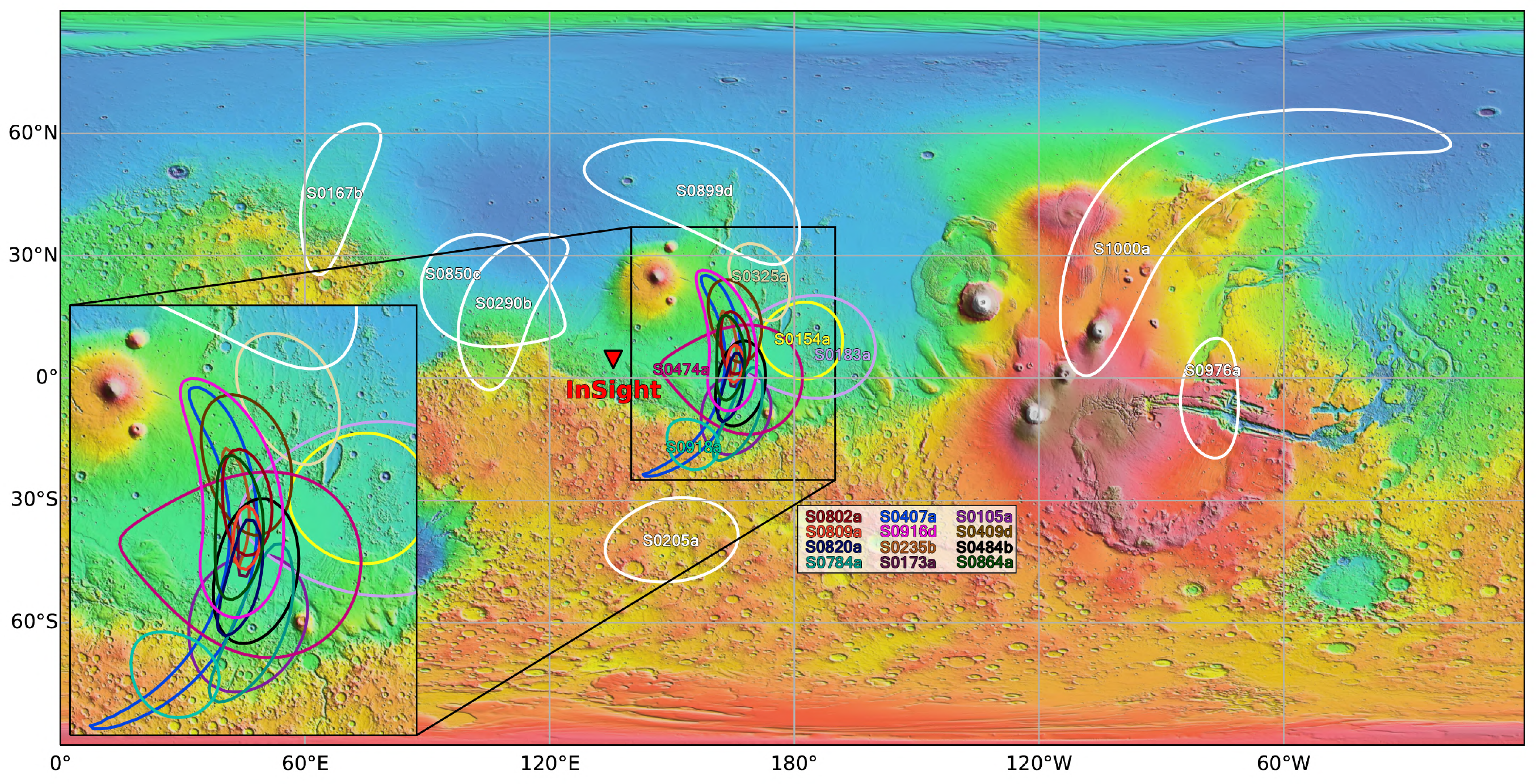}
\caption{Map of Mars using Mars Orbiter Laser Altimeter \citep[MOLA, ][]{smith_mars_2001} elevation data. The red triangle marks InSight's location in Elysium Planitia. The colored ellipses show the estimated locations of a set of high-quality marsquakes. Events close to the InSight lander are marked and labeled with colors, those further out are marked and labeled white. Distances with uncertainties are taken from the MQS catalog V9 \citep{MQSCatalog}. A zoom inset shows the area close to InSight where many quakes are located.}
\label{Fig:Map}
\end{figure}
\end{landscape}

\section{Seismicity of Mars}
The seismic data from Mars \citep{network_XB} is routinely analysed by MQS which catalogs any seismic activity \citep{MQSCatalog}. MQS categorises the seismic events recorded by InSight in different event families depending on their frequency content - low frequency families, and high frequency families. These two event families are then further divided into sub-categories. The low frequency family consists of low frequency (LF) and broadband (BB) events. LF events have energy below 1~Hz, whereas BB events have energy up to and around 2.4~Hz. This specific threshold is due to a subsurface structure effect resulting in a strong Airy phase at 2.4~Hz \citep{hobiger21}, which is visible in scattered energy excited by marsquakes \citep{clinton2021_catalog,Resonance2021}. The low frequency family are considered `classical' marsquakes, comparable to earthquakes. They occur at teleseismic distances and in depths of 20--50~km \citep{Brinkman2021, martian_core_2021} and have provided insight into the Martian interior.
The high frequency family consist of high frequency (HF), very high frequency (VF), and 2.4~Hz events. All three have energy predominately around 2.4~Hz and above. They are the more numerous group, with 882 out of 951 recorded events \citep{MQSCatalog}. In contrast to the low frequency family, their origin is less well constrained. So far, no back azimuth for a high frequency family event has been determined. Their predominantly high frequency content means that the signal is highly scattered and finding a stable back azimuth has proven challenging \citep{vanDriel2021_HF}. In this study, we focus on the LF and BB events recorded by InSight. For Catalog Version 9, only six out of the 69 recorded low frequency family events have been located with the standard MQS method: five in the general Cerberus Fossae region (events S0173a, S0235b, S0809a, S0820a, and S0864a), and one in the distant Valles Marineris region (event S0976a) \citep{MQSCatalog}. Two additional events have back azimuths but lack distances due to an uncertain (event S0183a) or absent (event S0899d) second phase. S0183a has a back azimuth consistent with Cerberus Fossae, but the second phase suggests the event is located further away in the Orcus Patera region. S0899d lies in northward direction, where the volcanic region of Elysium Mons can be found \citep{MQSCatalog}. \citet{khan2021upper} argue that the S-wave is not observed due to an S-wave shadow zone produced by a lithospheric negative gradient in $v_S$. Following that argument, the distance of these events would be in said shadow, between 45 and 55\textdegree.

\section{MQS back azimuth determination}
\begin{figure}[ht]
\centering
\noindent\includegraphics[width=0.8\textwidth]{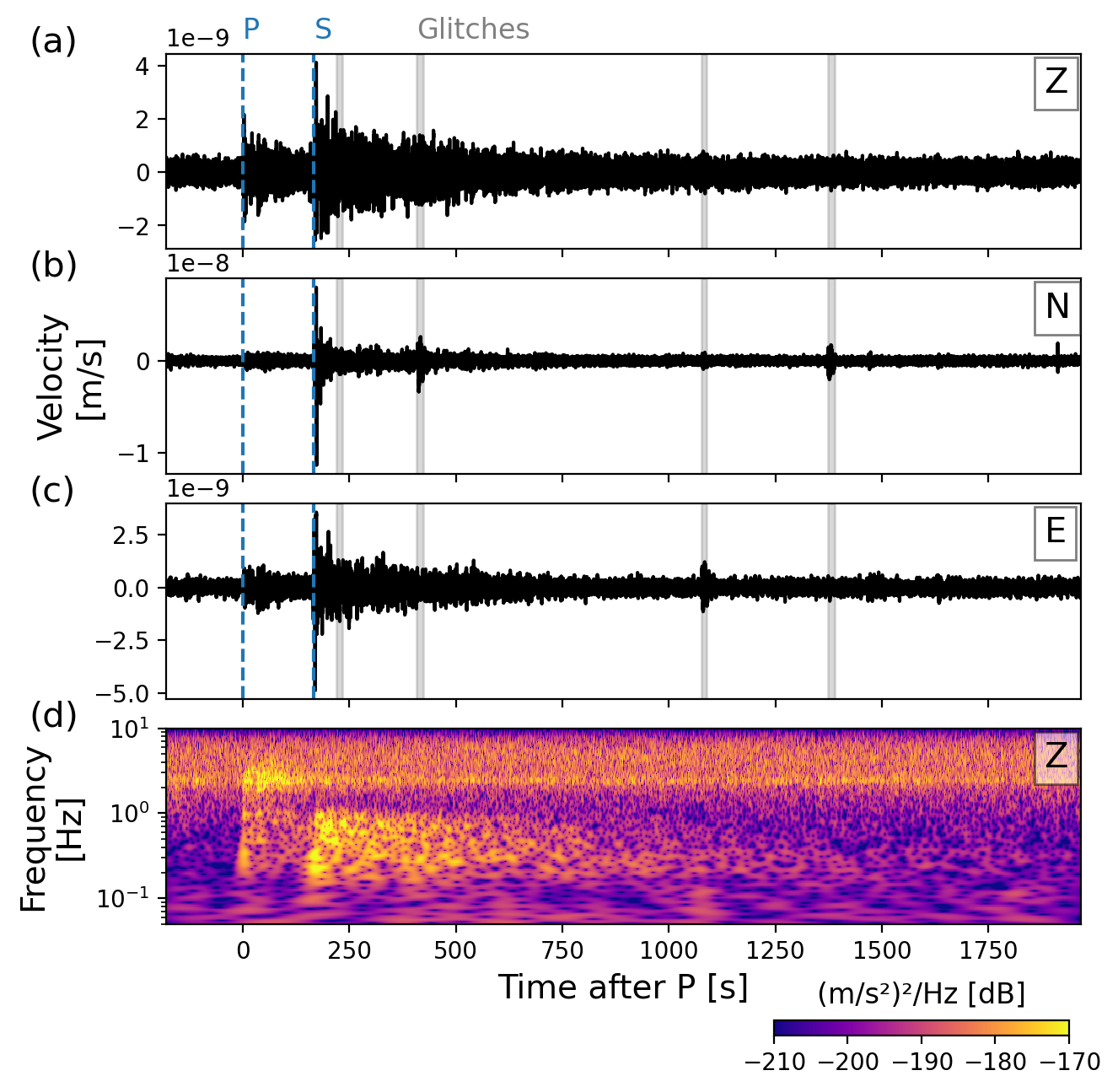}
\caption{Waveform timeseries of marsquake S0235b and the vertical acceleration spectrogram. Depicted are (a) the vertical (Z), (b) north (N), and (c) east (E) components of the velocity waveforms, 0.1~Hz--9.5~Hz bandpass. The MQS P and S arrival are indicated by the dashed blue lines. Glitches in the data are marked with gray boxes. (d) shows the vertical acceleration spectrogram between 0.05~Hz--10~Hz.}
\label{Fig:S0235b_waveform}
\end{figure}

Throughout the paper, for consistent presentation, we use the high quality low frequency family marsquake S0235b as an example. It is cataloged as a quality A broadband event with $M_w$ of 3.5 and a distance of 28$^{\circ}$. Its large signal to noise ratio (SNR) makes it one of the best events recorded by InSight so far. Figure \ref{Fig:S0235b_waveform} presents the waveforms for S0235b. Rows (a)-(c) show the velocity timeseries for the vertical, north, and east components respectively, filtered between 10~s and 9.5~Hz. The P and S arrivals identified by MQS are marked with a dashed line. Marked with gray bands are so-called \textit{glitches} in the data \citep{scholz20}, which we will explain in more detail later. Figure \ref{Fig:S0235b_waveform}(d) shows the acceleration spectrogram of the vertical component between 20~s and 10~Hz, calculated with a continuous wavelet transform. Both the P and S arrival are clearly visible in the spectrogram. The P-wave has more broadband energy and excites the 2.4~Hz resonance. On the other hand, the S-wave has lower frequency energy and is predominately visible below 1~Hz.

\begin{figure}[ht!]
\centering
\noindent\includegraphics[width=0.8\textwidth]{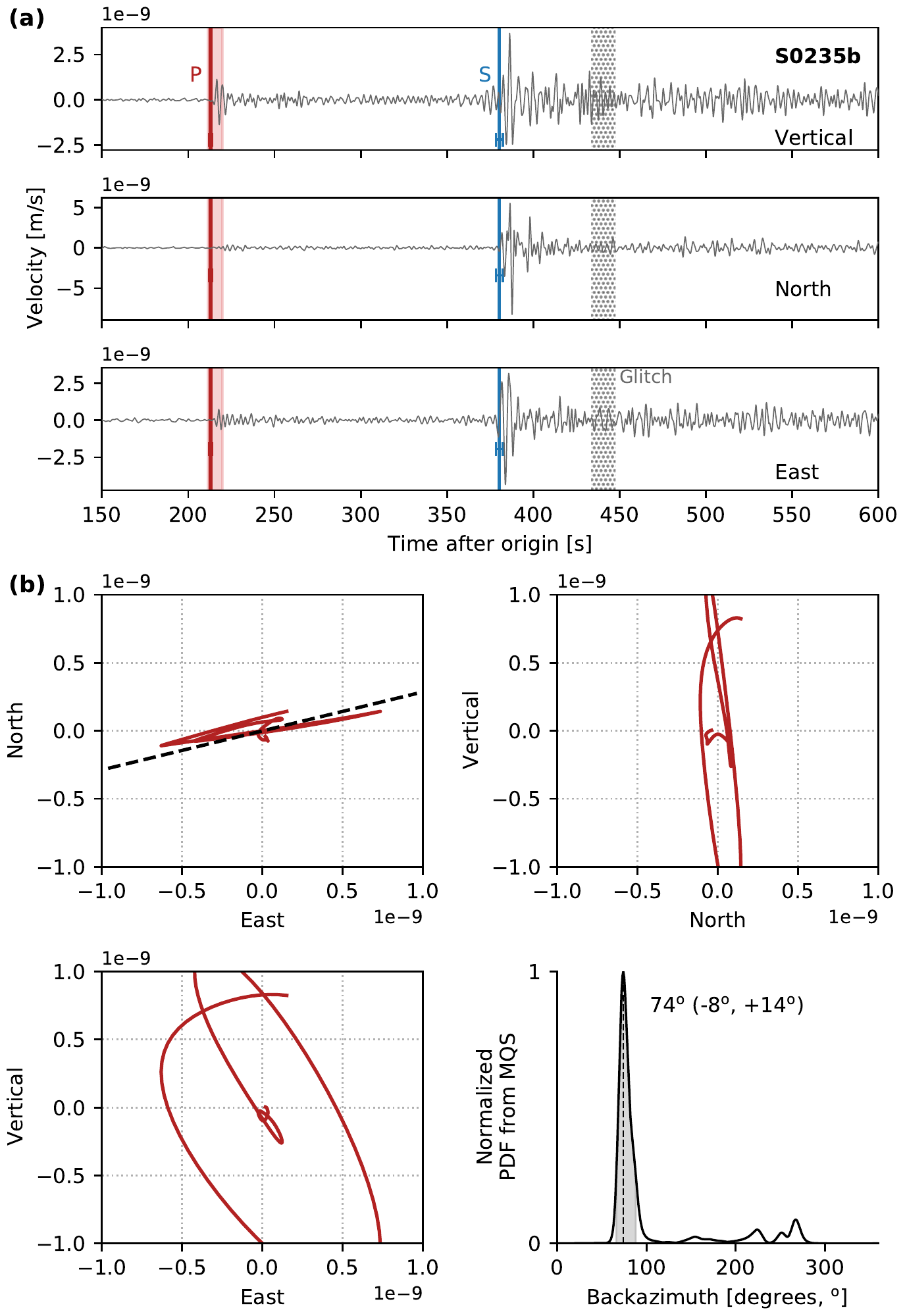}
\caption{MQS polarization analysis of marsquake S0235b. (a) Velocity waveforms for vertical, north, and east components filtered between 2--8~s. P (red) and S (blue) picks are marked with vertical lines. Picking uncertainties are given by a horizontal error bar. A glitch in the data is marked with the gray box. The light red box after the P pick marks the time window from which the hodograms in (b) are calculated. The normalized probability density function (PDF) is shown in the bottom right plot. The MQS-assigned back azimuth for this event is 74$^{\circ}$, indicated by the dashed black line.}
\label{Fig:hodogram}
\end{figure}
\begin{figure}[ht]
\centering
\noindent\includegraphics[width=0.8\textwidth]{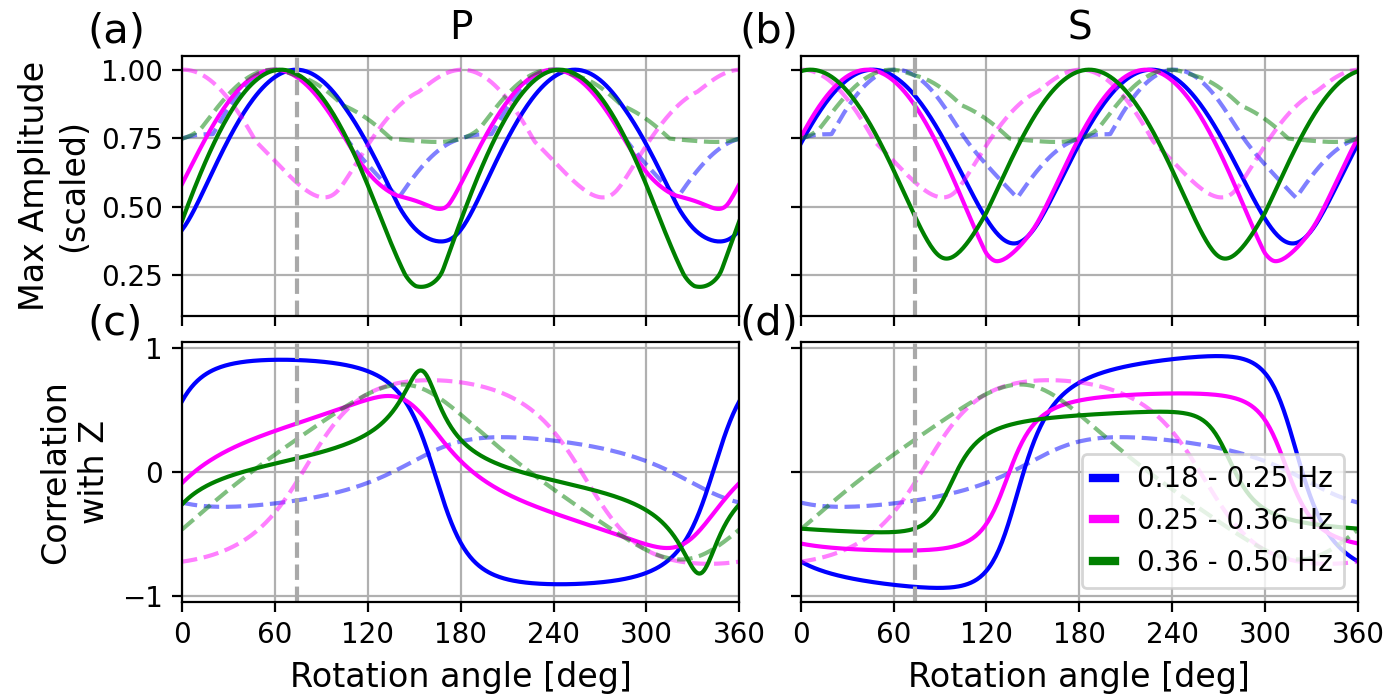}
\caption{Analysis of maximum amplitude of the rotated envelopes for marsquake S0235b. The normalised maximum amplitude of the radial component is given for a (a) P and (b) S time window. The correlation with the vertical component is given in (c) for P and (d) for S. The dashed lines in each plot show a pre-event noise window. Window lengths are each 15~s for the noise, P, and S window. The signal is bandpass-filtered for three frequency windows: 0.18--0.25~Hz (blue), 0.25--0.36~Hz (pink), 0.36--0.5~Hz (green).}
\label{Fig:S0235b_rotation}
\end{figure}

The method hitherto used by MQS to determine the back azimuth analyses the particle motion of a selected time window after the P-wave arrival, filtered in a frequency band that is optimized manually for each given event \citep{bose_probabilistic_2017}. The back azimuth $\Theta$ is determined from the amplitude ratio of the two horizontal velocity seismograms, here denoted by $e_i$ for east component and $n_i$ for the north component:
\begin{align}
    \theta_i = \begin{cases} 
    \arctan \left(e_i/n_i\right) & \textrm{if $z_i\cdot n_i < 0$} \\
    \arctan \left(e_i/n_i\right) + \pi & \textrm{otherwise} \\
    \end{cases} \label{Eq:theta}\\
    \Theta = \frac{\sum_{i=1}^{N} \theta_i \cdot (e_i^2+ n_i^2)}{ \sum_{i=1}^{N} (e_i^2+ n_i^2)},
    \label{Eq:mqs_baz}
\end{align}
where $i$ marks the sample index in the considered time window and $z_i$ the vertical component. The final back azimuth $\Theta$ is then obtained by averaging a weighted $\theta_i$ over the selected time window (Eq. \ref{Eq:mqs_baz}). The 180$^{\circ}$ ambiguity is resolved in Eq. \ref{Eq:theta} by enforcing the vector to lie in the lower hemisphere.
The polarization analysis of marsquake S0235b using this approach is summarised in Figure \ref{Fig:hodogram}. Figure \ref{Fig:hodogram} (a) shows filtered waveforms zoomed into the main body phase arrivals. The light red box around the P-arrival marks the time window from which the hodograms are calculated. As seen in these hodograms (Figure \ref{Fig:hodogram} (b)), the signal is predominately vertical although there is a consistent directionality in the horizontal plane. From the normalized probability density function, the back azimuth for S0235b is estimated as 74$^{\circ}$ by MQS. The assigned uncertainty is determined by the angles where the PDF amplitude falls to 25$\%$. Analysis of the hodograms works well for strongly polarized arrivals across broad frequency ranges. However, this method becomes unstable when particle motions are not persistent over the selected time window and a wide frequency range. This approach can also only be used for P-wave arrivals, and does not take advantage of the additional information that is contained in the S-wave arrival. Further, a number of marsquakes have been observed with high amplitude S-wave arrivals, but with very emergent or even undetectable P-waves. The choice of time window and bandpass filter can influence the result strongly for events of lesser quality. It is therefore difficult to obtain a robust result for the majority of the recorded events.

An alternative, simple method analyses the maximum amplitude of the seismic envelope of a time window after the P or S-wave. The seismic traces are rotated in small intervals over 360$^{\circ}$. The P-wave amplitude of the radial component is expected at a maximum when the rotation angle is the same as the back azimuth of the event. Conversely, the S-wave amplitude should be close to or at a minimum at the same time if assuming primarily SH-waves. There is a 180$^{\circ}$ ambiguity with this method, which can be resolved by calculating the correlation of the radial with the vertical component. The correlation is positive at the correct angle for the P-wave. There is also a sensitivity of the method to the frequency range of the bandpass filter applied to the seismic signal before rotating and calculating the envelope. The polarization of the signal can be (and routinely is) frequency dependent. We show three different frequency bands (0.18--0.25~Hz, 0.25--0.36~Hz, 0.36--0.5~Hz) for marsquake S0235b in Figure \ref{Fig:S0235b_rotation}. The MQS back azimuth is marked with a gray dashed line. In addition to a short 15~s window around the P and S-wave arrivals, we show a 15~s pre-event noise window to compare the signal to environmental noise. The P window maximum amplitude in Figure \ref{Fig:S0235b_rotation} (a) shows two maxima, around 60--70$^{\circ}$ and 240--250$^{\circ}$. The first coincides with a positive correlation with the vertical component (Fig. \ref{Fig:S0235b_rotation} (c)). This is in agreement with the results provided by MQS. The S window in Figure \ref{Fig:S0235b_rotation} (b) is more complex. In contrast to the P window, the frequency bands have less agreement and the minima diverge from the MQS results by about 60$^{\circ}$.
Different frequency bands only converge to a common back azimuth for high quality events. For all other events, the back azimuth from the maximum amplitude varies strongly between frequencies. This makes it impossible to say which, or if any, of the estimated back azimuths is correct. Both the MQS method and the rotated-amplitude method briefly presented here struggle with the vast majority of recorded marsquakes. Also, it would be preferable, if the tradeoffs in selection of time and frequency windows were immediately visible in the results.

In addition we compare our results to those from \citet{martian_core_2021} SM 4, currently under revision in \citet{Drilleau2021Locations}, who estimate back azimuths for a set of events based on Marsquake Catalog Version 8 (V8).
On top of the method described above, the instantaneous phase coherence between vertical and horizontal components as a function of azimuth. For P-waves, they expect the energy to be maximum and both correlation and phase coherence to be minimum at the correct angle. Since the interference of horizontal and vertical shear wave components (SH and SV, respectively) affects the energy, it cannot be used directly in the case of S-waves. However, the SV-wave component should produce a weak maximum of both correlation coefficient and phase coherence along the correct back azimuth. \citet{Drilleau2021Locations} analyse the signals in different time windows around their phase picks as well as a noise window. They use fixed frequency ranges for the analysis, 0.4--1.0~Hz for P-waves and 0.3--1.0~Hz for S-waves, owing to S-waves' lower frequency content. 

\section{Eigenvector method}
We propose to apply a complete polarization analysis of P and S body waves to determine the back azimuth of marsquakes.
The method is based on the work of \citet{Samson1980_polarization} and \citet{Samson1983_polarization}, and was further developed for seismological applications by \citet{Schimmel2003_polarization} and \citet{schimmel_polarized_2011} and has been applied to martian ambient noise before \citep{stutzmann2021polarized}. The 3-component seismogram is first transformed into time-frequency domain to produce a time-frequency dependent spectral matrix. For each time-frequency pixel, the matrix is decomposed into eigenvectors to obtain information on the polarization of the seismic signal. In contrast to \citet{stutzmann2021polarized}, we use the continuous wavelet transform as described by \citet{Kristekova_etal09} for the $t-f$ transformation.

\begin{figure}[ht]
\centering
\noindent\includegraphics[width=0.5\textwidth]{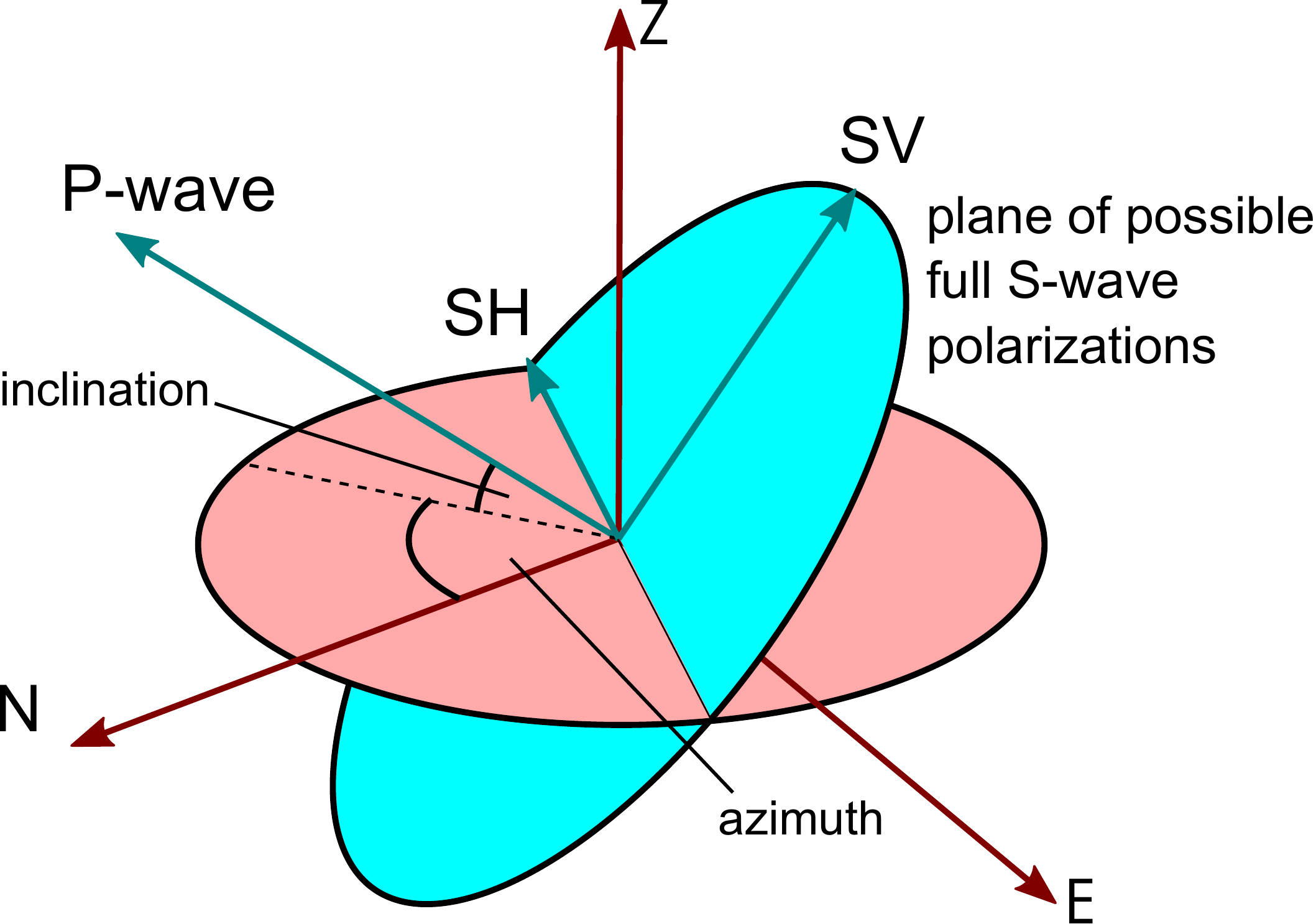}
\caption{Schematic showing wave polarization in 3-D space. The P-wave is represented by a vector with an inclination and azimuth relative to the horizontal plane (red ellipse) and north component (N), respectively. The S-wave consists of both SH- and SV-waves, which lie in a plane perpendicular to the P-wave vector (cyan ellipse).}
\label{Fig:S-polarization}
\end{figure}

Polarization describes the three-dimensional motion at the receiver location. At any particular time, this motion can be approximated by an ellipse with semi-major axis and semi-minor axis. From these axes in 3-D space, we can get the azimuth and inclination for each. We also obtain the ellipticity of the signal from the ratio of the semi-minor to the semi-major axis.
For an incoming P-wave, we expect the particle motion to be rectilinear and the azimuth of the semi-major axis of the polarization ellipse points in the direction of the event location \citep{sollberger2020}. The S-wave is more complex. To the first order and assuming an identical incidence vector between P and S, we expect the azimuth of the S-wave semi-major axis to be either $\pm 90^{\circ}$ (for SH) or $-180^{\circ}$ (for SV) with respect to the P-wave. The ratio of SH and SV-wave energy depends on the source mechanism. We propose an analysis that does not rely on correct discrimination between SH and SV-waves. Any incoming S-wave should lie in a plane perpendicular to the P-wave vector in 3-D space, see Figure \ref{Fig:S-polarization}. Assessing P and S-wave polarization together reduces the influence of noise in any back azimuth estimation.
Further, we can use the inclination of the signal to differentiate between the expected seismic signals and noise (e.g. wind or glitches in the data). For a P-wave, the inclination should be steep while a S-wave is expected to arrive with lower inclination.

The degree of polarization (DOP) gives a quality measure of the polarization based on the stability of the signal with time \citep{Schimmel2003_polarization,schimmel_polarized_2011}. A high-quality signal should not change its polarization during the signal, which forms the basis of the DOP. 
The DOP ranges between 0 and 1, where 1 describes a homogeneously polarized signal throughout the time window. It can be expressed in terms of eigenvalues $\lambda$ of the spectral matrix and using $n=3$ following \citet{Samson1980_polarization}
\begin{equation}
    \text{DOP} = \sum_{j,k=1}^n \left(\lambda_j - \lambda_k\right)^2 / \left[2(n-1) \left( \sum_{j=1}^n \lambda_j\right)^2 \right].
\label{Eq:dop}
\end{equation}

In this work, we adopt the following filter on the data to mask signals:

\begin{align}
\mathcal{F}_{DOP} = 
\begin{cases}
\text{DOP} \times 5 - 2, & \text{for } 0.4\leq \text{DOP}\leq 0.6\\
0, & \text{for } \text{DOP} < 0.4\\
1, & \text{for } \text{DOP} > 0.6
\end{cases}
\label{Eq:dop_filter}
\end{align}

where the filter $\mathcal{F}_{DOP}$ is linearly interpolated for DOP values in between 0.4 and 0.6. 

We further filter signals based on ellipticity, since we are only interested in rectilinear body wave arrivals. This filter is given by
\begin{equation}
\mathcal{F}_e = (1 - \epsilon)^{2 \alpha_{e}},
\label{Eq:elli_filter}
\end{equation}
where $\epsilon$ is the ellipticity of the signal with 0 being a rectilinear signal and 1 a circular signal, and $\alpha_e$ is a factor controlling the strength of the filtering. We adopt $\alpha_e = 1.0$ in this study to produce a low to moderate filtering effect. $\mathcal{F}_{DOP}$ and $\mathcal{F}_e$  are combined in an alpha filter

\begin{equation}
    \mathcal{F}_{total} = \mathcal{F}_{DOP} \times \mathcal{F}_{e},
\label{Eq:filter_tot}
\end{equation}
which is then applied to the data.
By selecting appropriate time and frequency windows for the P and S-wave arrivals, the polarization analysis presented is used to provide back azimuths for a large number of LF family marsquakes.

The most prominent noise source are wind gusts, which excite long period energy with predominantly horizontal polarization \citep{stutzmann2021polarized}. The aforementioned 2.4~Hz resonance is not strongly excited by local wind, but by marsquakes and thus visible for most quakes that have energy above the noise around 2~Hz. It is near-vertically polarized \citep{hobiger21}, with a roughly 180$^{\circ}$ azimuth \citep{Resonance2021}. The 2.4~Hz azimuth is independent of event polarization, but its exact azimuth seems to be methodology-dependent \citep{Resonance2021}. In addition to the 2.4~Hz resonance, there are a number of additional spectral peaks in the data. Below 9~Hz, there are resonances that are related to the measurement system (termed 1~Hz tick noise \citep{zweifel_seismic_2021} and its harmonic overtones) and to the lander itself (permanent lander modes at 1.6~Hz, 3.3~Hz, 4.1~Hz, 6.8~Hz, 8.6~Hz; transient lander modes at many other frequencies throughout the mission, \citet{Resonance2021}). Since these spectral peaks have their own stable azimuth, it is important to be aware of these and select frequency bands where they are not present. Though visible, the lander modes and 2.4~Hz resonance do not affect the analysis of the data since we focus on signal below 1~Hz in this study. The tick noise is removed in this analysis by averaging over a small frequency band around 1~Hz, so the influence of the 1~Hz tick noise is minimised. A further very significant source of polarized noise are the transient pulses that proliferate in the data termed glitches. They generally appear below 1~Hz with 25-30~s duration and have a strong and linearly horizontal polarization \citep{scholz20}. They are a major source of noise when analysing the polarization of a seismic signal, since even weak glitches show stronger polarization than seismic events. It is therefore important to select time windows that avoid glitches. We analyse the back azimuth in a select frequency band for each event, based on the event amplitude and P and S-wave inclination. Further, they are chosen such that they minimize the effect of any low frequency glitches and avoid known spectral peaks. By default, we also select a time window around the P and S picks identified by MQS from -5s to +10s. A noise time window is provided by MQS for each event.
We calculate the kernel density estimation \citep[KDE, ][]{scottKDE} across the specified time windows in the frequency band. The KDE curve maximum of the P-wave is the basis of the back azimuth determination. The full width at half maximum (FWHM) of the peak is calculated and provides a consistent uncertainty estimation. The KDE peaks of the S-wave are used to corroborate the results obtained from the P-wave. However, the S-wave is mainly assessed in the more complex 3-D analysis.

We estimate the P-wave vector from the peaks of the KDE curves of the azimuth and inclination. The S-wave vector should then lie in the plane perpendicular to this vector, and we can thus see whether the polarization contained in our S window agrees with our results from the P window. However, for weak or contaminated P-wave arrivals, we can turn this argument around and predict the P-vector from its orthogonality to the S-vector.
For any inclination-azimuth combination $(\theta,\phi)$, we calculate the cross-product with the polarization contained in the S window. Since the magnitude of the cross-product is maximum when the two vectors are perpendicular, this shows where the S window is perpendicular to - and thus, where we expect the back azimuth to be. This can be expressed as 
\begin{equation}
    \mathcal{L}(\theta,\phi) = \sum^{\theta'}\sum^{\phi'} \left | \vec{e}_P(\theta,\phi) \times \vec{e}_S(\theta',\phi') \right |,
\label{Eq:s-wave}
\end{equation}
where $\mathcal{L}$ is the P-vector 'likelihood', $\vec{e}_P$ the unit vector at location $(\theta,\phi)$ with length 1, and $\vec{e}_S$ is the vector at location $(\theta',\phi')$. Its length is given by the number of signals in the S window with $(\theta',\phi')$, so it is larger if more of the S window signal has a stable azimuth and inclination. Strongly polarized signals in the S window therefore affect the resulting 'likelihood' more strongly.
We can further combine this S-derived P-vector 'likelihood' with the information contained in the P window using
\begin{equation}
    \mathcal{L}_{P+S}(\theta,\phi) = \mathcal{L}(\theta,\phi) \mathcal{H}_P(\theta,\phi),
\label{Eq:P+S}
\end{equation}
where $\mathcal{H}_P$ is the 2-D histogram of the P window in the selected frequency range for inclination and azimuth $(\theta,\phi)$. We add a 'water level' to $\mathcal{H}_P$ to prevent it from dominating the result, since a large part of the inclination-azimuth space of $\mathcal{H}_P$ is zero.

\section{Verification using synthetic marsquakes}
We first test our method on a set of synthetic marsquakes.
For a best-case comparison with real marsquake data, we use noise recorded from a quiet evening without marsquakes and add a synthetic event to the data. This noise data includes the typical signals (e.g. glitches, tick noise), that are thus handled the same as real marsquakes. Sol 245 (a sol is a Martian day) has low noise conditions and several hours of quiet data available. We use the Mars model \textit{InSight\_KKS21GP} \citep{KKS_2021GP} to generate synthetic waveform data (see Data section). We use a frequency band between 0.3--1.0~Hz for the determination of the back azimuth, as this catches the main energy of the synthetic event while avoiding noise contamination at low frequencies. Since this is a 1-D model, the position of source and receiver is not relevant and only distances, magnitude, and source depth are relevant factors. The source is placed to the east of the receiver at 50~km depth, making the true back azimuth (BAZ$_{true}$) 90$^{\circ}$. In the example presented in this section, the distance is 40$^{\circ}$. We use a focal mechanism of strike = 280$^{\circ}$, dip = 79$^{\circ}$, and rake = -79$^{\circ}$, similar of that estimated for marsquake S02035b by \citet{Brinkman2021}. The moment magnitude $M_w$ is set to 3.5, which is expected for medium to high quality events. The observed marsquake data differ from these synthetic seismograms mainly by strong scattering, which decreases the amount of polarization in the body wave coda and suppresses surface waves. For the initial arrival window of body waves, these parameters should make this synthetic event comparable to high quality marsquakes.

\begin{figure}[htb!]
\centering
\noindent\includegraphics[width=0.85\textwidth]{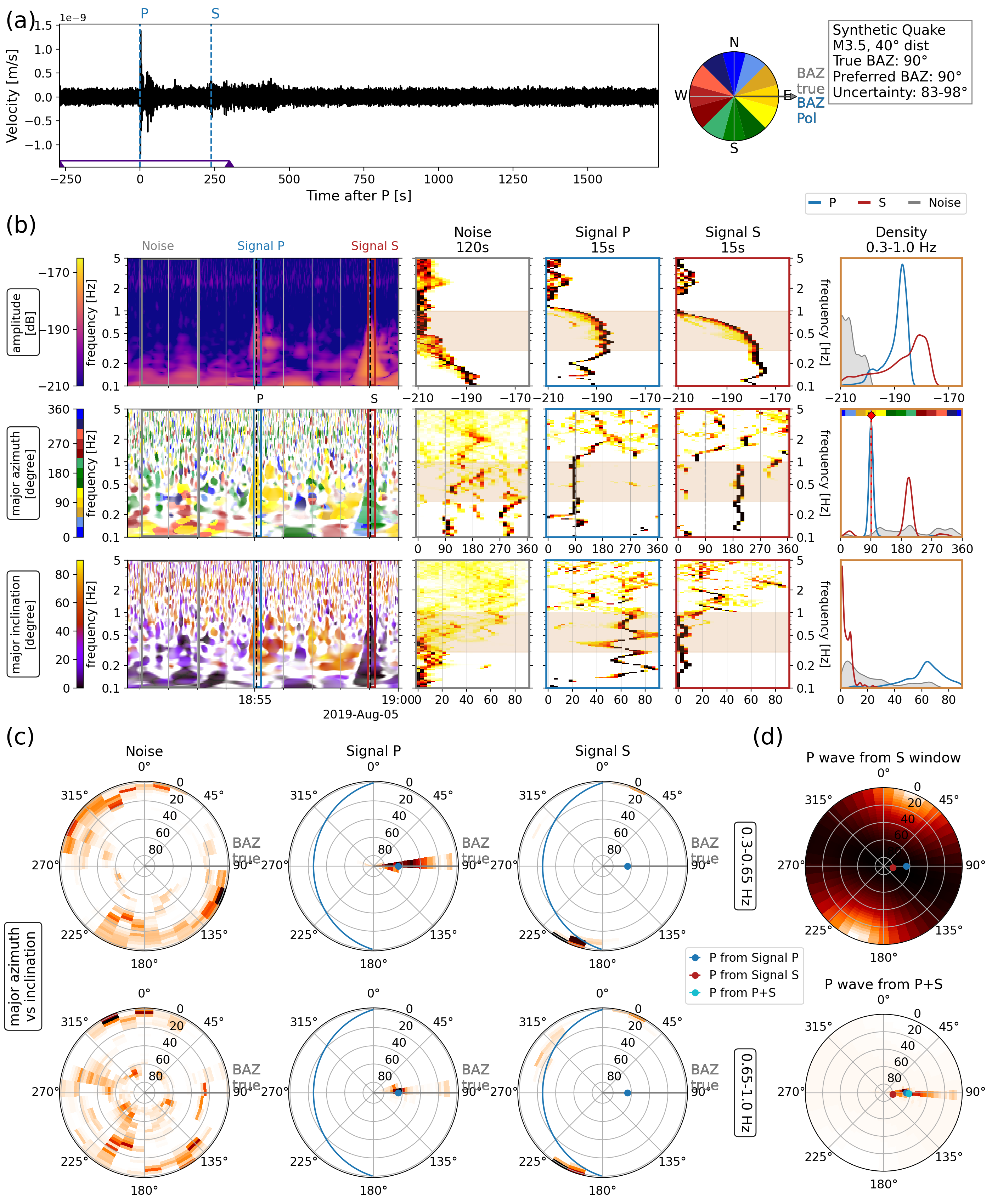}
\caption{Polarization analysis of a synthetic marsquake with $M_w~=~3.5$ and a distance of 40$^{\circ}$. (a) Vertical velocity waveforms with P and S arrivals (blue dashed lines). (b) Polarization analysis based on the velocity timeseries, (b,top) amplitude in $(m/s)^2/Hz$ [dB], (b,middle) azimuth, (b,bottom) and inclination. (b,left) time-frequency plots for each parameters, with three marked windows: pre-event noise (gray),  P-wave arrival (blue), S-wave arrival (red). Phase arrival signal windows are -5~s to 10~s around the corresponding pick, and the noise window has 120~s duration. (b,columns 2--4) Histogram representations of each of the three windows marked in the left column, with the amplitudes on the x-axis corresponding to the scale shown in the time-frequency colorbars from the left column. Vertical dashed gray lines mark the true back azimuth. Shaded areas indicate the preferred frequency band (0.3--1.0~Hz). (b,right) KDEs calculated within the shaded frequency band for each time window. The back azimuth estimated from this analysis is defined as the peak of the P-wave azimuth. Dashed gray and solid red lines in the second row mark the true and calculated back azimuth, respectively. 
(c) Azimuth-inclination histograms for the time windows separated into two frequency bands (0.3--0.65~Hz, 0.65--1.0~Hz). KDE-derived P-wave vector and its perpendicular plane are marked by a blue dot and curve, respectively. (d) P-vector estimation from the S window (top), and combination of P and S window results (bottom). The S-derived P-vector is marked with a red dot.}
\label{Fig:Syn_polarization}
\end{figure}

The polarization analysis for the synthetic marsquake is seen in Figure \ref{Fig:Syn_polarization}. In Figure \ref{Fig:Syn_polarization}(a), we show the vertical velocity waveforms with P and S arrivals (blue dashed lines), bandpass filtered 0.3--1.0~Hz (the same frequency band used for the back azimuth estimation described later). The S-wave arrival is only weakly visible on this component. The purple triangles and line mark the time window which is shown in (b). We first focus on the time-frequency depiction of the polarization analysis (Fig. \ref{Fig:Syn_polarization}(b)). The data are shown between 0.1--5~Hz. Each row shows different parameters: the first row shows the spectral amplitude of the signal in units of (m/s)$^2$/Hz [dB], combined from three components. The middle row shows the azimuth of the semi-major axis of the polarization ellipse, and the last row shows the inclination of the semi-major axis. The leftmost column shows a time-frequency representation of the data. All parameters are filtered to enhance polarized signals in the plot. For rows two and three, non-polarized signals are masked with the alpha filter from Eq. \ref{Eq:filter_tot} and appear white. Three time-frequency windows are marked with boxes: a 2~min pre-event noise window (gray), a 15~s P arrival window (blue), and a 15~s S arrival window (red). The two signal windows are from -5~s to +10~s around the known phase arrivals (data before the arrival is used since the continuous wavelet transform is acausal). The amplitude / frequency information from these 3 time windows are shown in the second, third, and fourth columns. The amplitudes in the x-axes correspond to the colorbar values in the time-frequency plots. The vertical gray dashed lines in the middle row mark the known BAZ$_{true}$ for the synthetic event.
The beige shaded part in the histograms mark the frequency band used for the KDE shown in the rightmost column. The maximum of the P window KDE curve is marked with a solid red line. The compass rose at the top of the figure shows how the azimuth colours are related to geographic directions. The gray arrow marks the direction of BAZ$_{true}$, the blue arrow marks the direction of the polarization-derived back azimuth (BAZ$_{pol}$).

The event is clearly visible at frequencies below 1~Hz. The maximum amplitude is -185~dB for the P and -175~dB for the S-wave. The noise window has amplitudes between -200 to -185~dB, with higher amplitudes at lower frequencies. In the preferred target frequency band we use to determine the back azimuth, between  0.3--1.0~Hz, the noise window has amplitudes more than 20~dB - 1 order of magnitude - below that of the signal windows. The SNR of the P window between 0.3--1.0~Hz is 6.2. When taking a closer look at the time-frequency plots in rows 2--3, parts of the signal have been filtered out. At 18:52:30~UTC, there is a small glitch visible in the noise window with east-west polarization. The P and S window are strongly polarized up to 1~Hz. The P-wave polarization persists for around 1~minute after the arrival of the signal. The S-wave polarization is shorter and more focused around the arrival of the wave. The back azimuth estimated from the KDE of the P-wave is 90$^{\circ}$, equal to BAZ$_{true}$, with an uncertainty of 83--98$^{\circ}$. The KDE maximum of the S-wave is around 200$^{\circ}$, roughly orthogonal to the P-wave. The noise window has no clear peak or preferred azimuth. As expected for an event at this distance with near vertically incident body wave arrivals, the P-wave has a steep inclination of 50--80$^{\circ}$, which contrasts  with the S-wave with an inclination close to 0$^{\circ}$.

In summary: Figure \ref{Fig:Syn_polarization}(b) shows that there is a clear difference in polarization between the pre-event noise window, the P-wave arrival, and the S-wave arrival. The properties are as expected: the P-wave has a high inclination, rectilinear signal in direction of BAZ$_{true}$; the S-wave has a low inclination, rectilinear signal with shifted azimuth. The noise window is not strongly polarized and has no preferred azimuth as described by \citet{stutzmann2021polarized} - though some noise features of the InSight dataset are known exceptions, including glitches. The back azimuth of the event estimated from the P window matches BAZ$_{true}$.

The azimuth of the three selected time windows (noise, P, and S) is analysed in more detail in Figure \ref{Fig:Syn_polarization}(c).
The top and bottom rows show histograms of the semi-major azimuth on the angular axis against the inclination on the radial axis. They show the data for a lower (0.3--0.65~Hz, top row) and higher (0.65--1.0~Hz, bottom row) frequency band. To mimic a stereographic projection, the inclination is 90$^{\circ}$ in the middle of the plot and goes to 0$^{\circ}$ at the outer edge. We see that the noise window (left) has a low inclination and a wide range of azimuths, consistent with wind-induced noise \citep{stutzmann2021polarized}.
To help analyse the seismic phases, we mark the P-wave vector in the P and S window plots. It is described by an azimuth and an inclination, which are taken from the maxima of the respective KDE in Figure \ref{Fig:Syn_polarization}(b). The vector's intersection with the half-sphere reduces it to a dot. We expect the S-wave to be perpendicular in 3-D space to this vector, i.e. located on a plane with the P-vector as the normal vector. Thus, we calculate the plane perpendicular to the vector and its intersection with the half-sphere, which results in a curve. These two features are plotted in blue in Figure \ref{Fig:Syn_polarization}(c). For the middle column (corresponding to the P window), we expect the data to cluster around the blue dot. For the right column (corresponding to the S window) the data should cluster somewhere on the blue curve.
The data is clustered around the P vector dot in the middle column as expected. The right column shows three clusters, south, north, and northwest, with very low inclination visible in both frequency bands. They lie close to the blue curve and are thus perpendicular to the estimated P-vector. They agree with our predicted S-wave and we consequently interpret these as the S-wave.

Figure \ref{Fig:Syn_polarization}(d) shows the predicted P-wave based solely on the polarization contained in the S window. The background shading is calculated using Eg. \ref{Eq:s-wave}, where dark areas have a higher 'likelihood' for the P-vector, i.e. larger parts of the S window polarization are perpendicular to these areas. The maximum is marked by a red dot, giving the P-vector as predicted by the S window polarization alone. For comparison, the P window KDE-derived P-vector is given by the blue dot. We see in Figure \ref{Fig:Syn_polarization}(d,top) that these two dots agree well with each other, and the main difference is the inferred inclination of the P-vector. Figure \ref{Fig:Syn_polarization}(d,bottom) combines the information from the P and S windows by multiplying the azimuth-inclination histogram of P between 0.3--1.0~Hz (Fig. \ref{Fig:Syn_polarization}(c) combined middle column) with the background from Figure \ref{Fig:Syn_polarization}(d,top) as described in Eq. \ref{Eq:P+S}. Marked in cyan is the resulting P-vector. Since these two results are so similar, combining them to obtain a third set of P-vector offers little improvement for this synthetic case.

\section{Verification using earthquakes}
\begin{figure}[ht!]
\centering
\noindent\includegraphics[width=\textwidth]{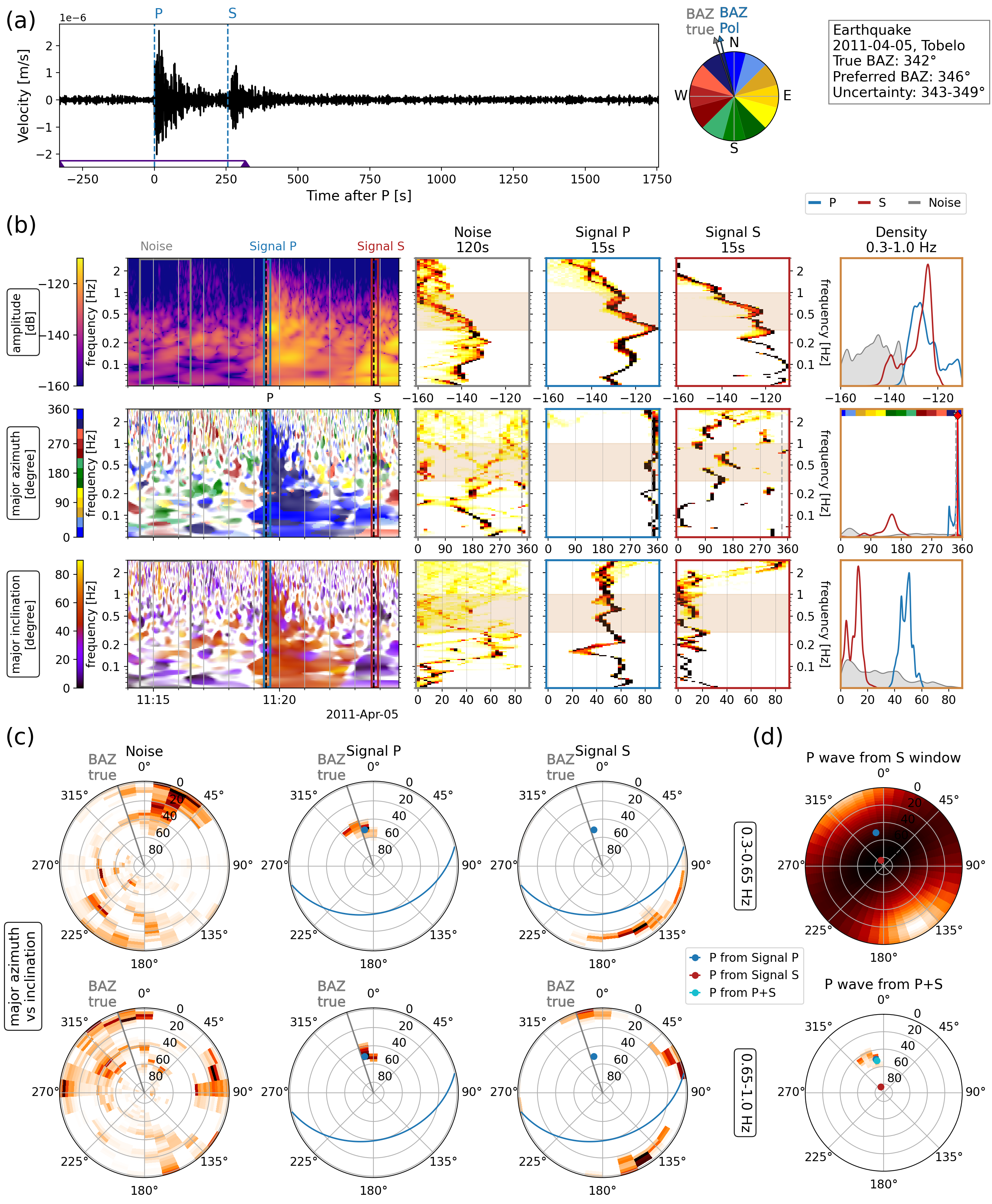}
\caption{Polarization analysis for an earthquake recorded at station WRAB, Australia on April 5th, 2011. The plot follows the same structure as Fig. \ref{Fig:Syn_polarization}. The earthquake has $M_w=5.9$, and the back azimuth is 342$^{\circ}$.}
\label{Fig:Earthquake_polarization}
\end{figure}

To test the method on events with known back azimuth, we take a dataset recorded at the Warramunga array, station WRAB, located in Tennant Creek, Australia, which has similar topographic conditions as InSight on Mars (i.e. flat surroundings and no close-by mountains) \citep{network_II}. Similar to the results in the previous section, we look at the polarization of an earthquake which occurred in Indonesia on April 5th, 2011. The source was at 3.03$^{\circ}$ N 126.94$^{\circ}$ E, in 20~km depth with $M_w = 5.9$ \citep{USGS_Tobelo} at an epicentral distance of 2666~km.
Figure \ref{Fig:Earthquake_polarization} shows the polarization analysis with the same format as shown in Figure \ref{Fig:Syn_polarization}. The event is clearly visible on the vertical waveforms (Fig. \ref{Fig:Earthquake_polarization}(a)). There is excess energy up to 3~Hz for the P-wave visible in the spectrogram (Fig. \ref{Fig:Earthquake_polarization} (b,top), first column). The S-wave is visible at lower frequencies and has strong excess energy around 0.1~Hz. The amplitude KDE shows strong excess energy for both phases, with an SNR for the P-wave of 48.5.
When considering the polarization-filtered azimuth in the second row, the event is clearly visible in the time-frequency plot on the left. The P-wave in particular has a strong, consistent polarization which differs from the noise. The S-wave polarization appears slightly weaker, and shows no dominant azimuth above 1~Hz. The back azimuth estimated from the KDE (right column) of the P-wave is 346$^{\circ}$ with an uncertainty of $343\textbf{} - 349^{\circ}$. The estimated back azimuth is very close to the true back azimuth of 342$^{\circ}$. The S window has several peaks, roughly along and opposite to the true back azimuth. The noise window is differently polarized from the earthquake signal.
The P window has an (primarily moderate to high) inclination of 40 to 60$^{\circ}$. The S window has a low inclination between 0--20$^{\circ}$. In contrast, the noise window has a wider distribution with less prominent singular peaks.
Figure \ref{Fig:Earthquake_polarization}(c) provides a deeper analysis of the azimuth of the time windows, where as before, the plot shows azimuth against inclination for two frequency bands. The true back azimuth is marked with a gray line. 
The noise window has a distinctly different polarization from the P and S windows. The P window matches the true back azimuth (gray line) well in both frequency bands, and consequently,  the P vector (blue dot) estimated from the polarization analysis lies very close to the true back azimuth. 
Several clusters in the S window lie close the blue curve, meaning they are orthogonal to the P-wave in 3-D space. It appears that the KDE-derived P-vector is too low (around 50\textdegree) for an optimal fit with the S window polarization. A more vertical inclination for the P would shift the blue curve towards lower inclination, improving agreement with the S window. The fit with the expected orthogonal constraint is good in both frequency bands. In the higher frequency band, there is some additional horizontal energy along the back azimuth direction.
Figure \ref{Fig:Earthquake_polarization}(d,top) shows the estimated back azimuth from the S window. The P-vector inclination expected from the S window is much steeper than estimated from the P window itself, but the back azimuth is consistent between both. Combining the P and S window (Fig. \ref{Fig:Earthquake_polarization}(d,bottom)) gives a result very similar to that of the P window, and thus close to the true back azimuth.

\begin{figure}[ht]
\centering
\noindent\includegraphics[width=0.8\textwidth]{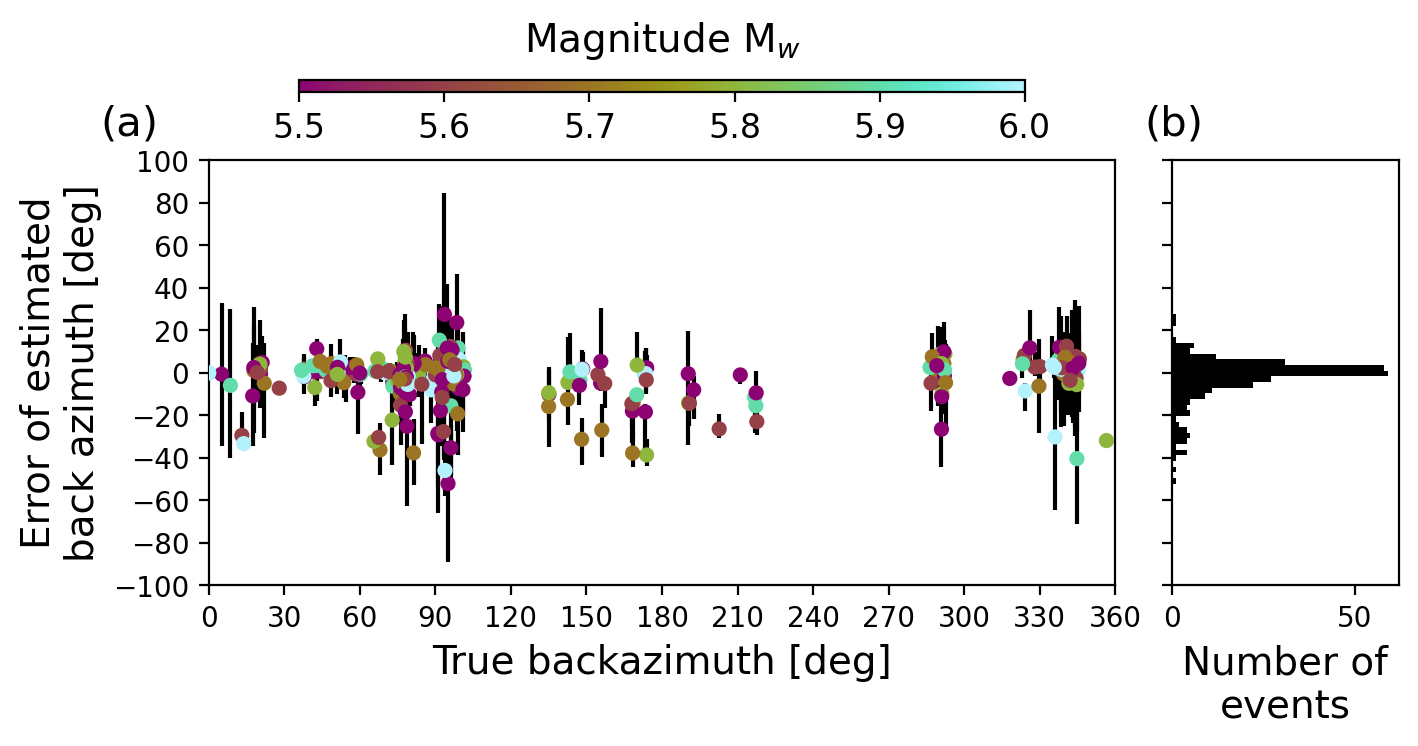}
\caption{Error of polarization-derived back azimuth for earthquakes recorded at station WRAB, Australia between 2010 and 2017. Shown are (a) the error in back azimuth versus the true back azimuth, and (b) a histogram of the error distribution. Also depicted are the earthquake magnitude $M_w$ (color dots) and estimated uncertainty (black lines). $M_w$ is between 5.5 and 6.0 for all events, resulting in a total of 264 events.}
\label{Fig:BAZ_error}
\end{figure}

While the analysis of single events is useful for comparison with marsquakes, in particular regarding the combination of inclination and azimuth, a more systematic approach is needed to assess how the method performs on a larger scale.
We analyse a set of 264 earthquakes recorded on the same seismic station (WRAB) to assess the error in estimated back azimuth. Systematic errors could be related to a local site effect. The set of earthquakes has distances between 20$^{\circ}$ and 40$^{\circ}$, source depths of 10 to 30~km, $M_w$ of 5.5 to 6.0, and occurred between beginning of 2010 and end of 2017. These parameters should make the results comparable to Mars (i.e. adjusted magnitudes to account for Earth's higher noise level, relatively shallow sources) and the epicentral distance covers several seismically active regions from different directions. We use the maximum of the P window KDE to estimate the back azimuth, and we estimate the uncertainty using the peak-half-width. All earthquakes are analysed between 0.3--1.0~Hz. Figure \ref{Fig:BAZ_error} (a) shows the true back azimuth BAZ$_{true}$ versus the polarization-derived back azimuth BAZ$_{pol}$. The error is calculated as $\Delta $BAZ $=$ BAZ$_{true} - $BAZ$_{pol}$, so positive (negative) $\Delta$BAZ values correspond to too small (large) estimated BAZ values. The black bars show the estimated uncertainty. Earthquake magnitude is given by the marker color. A histogram of the back azimuth error is shown in Figure \ref{Fig:BAZ_error} (b).
The polarization analysis matches BAZ$_{true}$ well, with an error of less than 10$^{\circ}$ in 213 out of 264 earthquakes. Only 13 events have an error larger than 30$^{\circ}$. There is no clear trend in the error depending on the back azimuth of the earthquakes, so no obvious site effect is visible in this analysis. The distribution of earthquake back azimuths is, as is expected, not uniform but clustered, related to active fault boundaries near Australia.

\section{Application to Marsquakes}
\subsection{Back azimuth determination for selected marsquakes}
\begin{figure}[ht!]
\centering
\noindent\includegraphics[width=\textwidth]{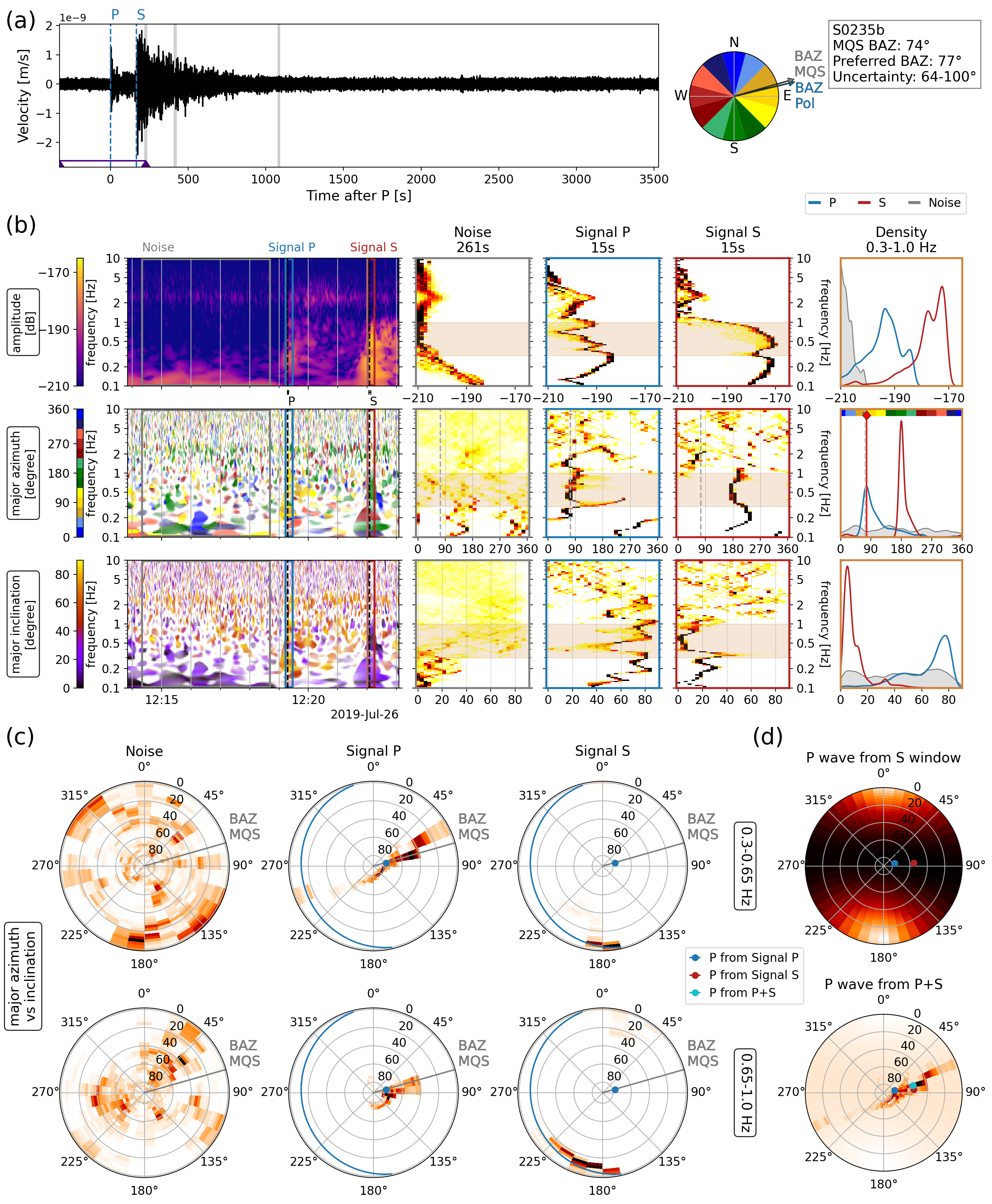}
\caption{Polarization analysis for marsquake S0235b ($M_w$3.7 BB QA event at 27.9$^{\circ}$ distance, July 26, 2019). The plot follows the same structure as Fig. \ref{Fig:Syn_polarization}. The signal windows are -5~s to 10~s around the corresponding wave pick. The noise window is used from MQS, with 261~s length. Pick uncertainties are taken from MQS and are indicated by horizontal bars (b, top row, left column). The MQS back azimuth is marked with a gray dashed line in the second row in (b), and with a gray line in (c). The KDE maximum of the P window in (b, right column) is marked with a red line and diamond indicator. The 1~Hz tick noise has been reduced to avoid contamination of the polarization analysis.}
\label{Fig:S0235b_polarization}
\end{figure}

After testing the method on synthetic marsquakes and a number of earthquakes, we apply the polarization analysis to a set of LF and BB marsquakes.
The polarization analysis for marsquake S0235b is seen in Figure \ref{Fig:S0235b_polarization}. Same as before, the vertical velocity waveforms are shown bandpass filtered between 0.3--1.0~Hz (Fig. \ref{Fig:S0235b_polarization}(a)). In (b), we show the time-frequency polarization analysis. The amplitude spectrogram in the top left is in velocity, in contrast to the acceleration spectrogram shown in Figure \ref{Fig:S0235b_waveform}. In addition to the P and S-waves, the 2.4~Hz resonance is visible with roughly southward azimuth and near-vertical inclination.
Unlike many other marsquakes, S0235b has no glitches close to either the P or S pick. The top row in (b) shows the large signal to noise ratio of the event. Therefore, the polarization for the P and S window should be solely due to the event and not be dependent on background noise. The KDEs for the azimuth ((b), right column, second row) show a distinct peak for both P and S windows, and a flatter distribution for the noise window. The three time windows differ strongly in inclination: the S window has a very low inclination while the P as a steeper inclination. The noise has a broader distribution.
The estimated back azimuth from the polarization matches the one from the MQS catalog \citep[gray dashed line, ][]{MQSCatalog}.
We examine the three time windows more closely in Figure \ref{Fig:S0235b_polarization}(c). The MQS back azimuth is marked with a gray line. The noise shows no preferred azimuth between 0.3--1.0~Hz (Fig. \ref{Fig:S0235b_polarization}(c), left column).
The P window shows a clear preferred azimuth over the whole analyzed frequency band. The P window signal has a steep inclination, best visible between 0.65 and 1~Hz. At lower frequencies, the inclination also contains some more horizontal polarization.
The S window has two discernible azimuths, roughly southwards and, less dominant, to the north-east. The latter is weakly visible in the higher frequency band, while the former is visible over the whole frequency range. The S window azimuth-inclination signal lies clearly on the blue curve, perpendicular to the estimated P-vector (blue dot).
Figure \ref{Fig:S0235b_polarization}(d) shows the expected P-vector as determined solely from the S window, and compares it with the P window-derived P-vector. While inclinations differ slightly, the back azimuth is consistent between both approaches.

We apply the same methodology to a number of other marsquakes, all of the LF or BB event category \citep{clinton2021_catalog}, which are selected for clear phase arrivals. We estimate the back azimuth from the maximum of the P-wave, and corroborate this with the S-wave, whenever possible. The back azimuth uncertainty is determined by the half-width of the P-wave KDE. Results are summarised in Table \ref{Tab:Results_marsquakes} and Fig. \ref{Fig:results_pdf}. A detailed polarization analysis of each event in the table can be found in the Supplementary Material together with a short description. The upper part of Table \ref{Tab:Results_marsquakes} shows results for the best marsquakes with clear phase arrivals and strong polarization. The lower part of the table shows results for events with less clear polarization. They often require a combination of P and S-wave polarization analysis, since the P-wave polarization alone is not sufficient, it is contaminated by significant glitches, or the phase arrival timing is uncertain. Results are therefore less certain. We find that while events with large SNR typically mean reliable back azimuths can be estimated,  events with smaller SNRs can still provide back azimuths if the polarization of the event is sufficient. The SNR can also be affected by a glitch close the P-wave onset, or by phase picks with large uncertainty.
We see that results obtained from the polarization analysis are close to all those provided by MQS, where available, but that we do find new back azimuths for 16 more quakes.

\begin{landscape}
\begin{table}
\centering
\caption{Table summarising marsquake back azimuth estimates. The marsquake events, type, quality, and MQS back azimuth are taken from the MQS catalog \citep{MQSCatalog}. MQS uncertainties are described in \citet{clinton2021_catalog}. Results from \citet{Drilleau2021Locations} are given in 'Drilleau (2022)'. SNR is calculated from the P-wave and indicated noise window in the same frequency band where the back azimuth is estimated. Uncertainty for the polarization back azimuth is calculated from the width of the KDE at half-maximum. Events in \textit{italics} have less clear results and are more speculative.}
\begin{tabular}{llllllllll}
\toprule
\multirow{2}{*}{Event} & \multirow{2}{*}{Type} & \multirow{2}{*}{Quality} & \multirow{2}{*}{SNR$_P$} & \multicolumn{6}{c}{Back Azimuth [deg]}\\ 
\cmidrule(lr){5-10}
                             &                                        &  &  & MQS  & Uncertainty & Drilleau (2022) & Uncertainty & Polarization  & Uncertainty    \\
\midrule
S0173a & LF   & A & 26.6       & 91  & 79-102   & 88.2 &72.2-104.2  & 88      & 78-103    \\
S0183a & LF   & B & 9.4       & 73  & 61-82  & - &  - & 85      & 67-101   \\
S0235b & BB   & A & 37.2       & 74 & 66-88  & 69 & 50.9-87.1 & 77      & 64-100    \\
S0802a & BB   & B & 6.5       & -  & -    & 85 & 65.5-105 & 82      & 65-96   \\
S0809a & LF   & A & 15.9       & 87 & 67-105   & 86.3 & 71-101  & 91      & 82-100    \\
S0820a & LF   & A & 12.0       & 88  & 76-107  & 84 & 66.1-102  & 106      & 85-120    \\
S0864a & BB   & A & 2.3       & 97  & 83-116  & 88 & 61-115 & 90      & 66-110   \\
S0899d & LF   & B & 35.7       & 25 & 11-37   & - & - & 22      & 354-55    \\
S0916d & BB   & B & 3.5       & -  & - & 71 & 45.3-96.3 & 97      & 41-114    \\
S0976a & LF   & A & 35.7       & 101 & 71-121   & - & - & 94      & 65-119    \\
\midrule
\textit{S0105a} & LF   & B & 5.1       & -  & - & - & - & 112     & 95-133   \\ 
\textit{S0154a} & BB   & C & 5.5       & -  & - & 87 & 31.4-142.6 & 82      & 69-95    \\
\textit{S0167b} & LF   & C & 2.5       & -  & - & - & - & 315      & 295-335$^\dagger$    \\
\textit{S0205a} & BB   & C & 4.0       & -  & - & - & - & 164      & 147-182    \\
\textit{S0290b} & LF   & B & 1.5       & -  & - & - & - & 292      & 257-341    \\
\textit{S0325a} & LF   & B & 14.1       & -  & - & 126 & 109-143 & 57      & 43-73    \\ 
\textit{S0407a} & LF   & B & 3.4       & -  & - & 79 & 54.5-104 & 57      & 43-169    \\ 
\textit{S0409d} & LF   & B & 4.0       & -  & - & 82 & 57-107 & 70      & 50-90$^\dagger$     \\
\textit{S0474a} & LF   & C & 0.9       & -  & - & 21 & 332-69.8 & 97      & 72-123    \\
\textit{S0484b} & BB   & B & 3.4       & -  & - & 73 & 39.3-107 & 100      & 80-120$^\dagger$    \\
\textit{S0784a} & BB   & B & 8.4       & -  & - & 101 & 83.5-118 & 115     & 92-136    \\ 
\textit{S0850c} & BB   & C & 1.4       & -  & - & - & - & 301      & 277-325    \\
\textit{S0918a} & LF   & B & 5.7       & -  & - & 162 & 70.6-252 & 137      & 125-149    \\
\textit{S1000a} & BB   & B & 21.6       & -  & - & - & - & 57      & 345-87    \\
\bottomrule
\multicolumn2l{$^\dagger$ Uncertainty set manually.}
\end{tabular}
\label{Tab:Results_marsquakes}
\end{table}
\end{landscape}

\begin{figure}[ht!]
\centering
\noindent\includegraphics[width=0.8\textwidth]{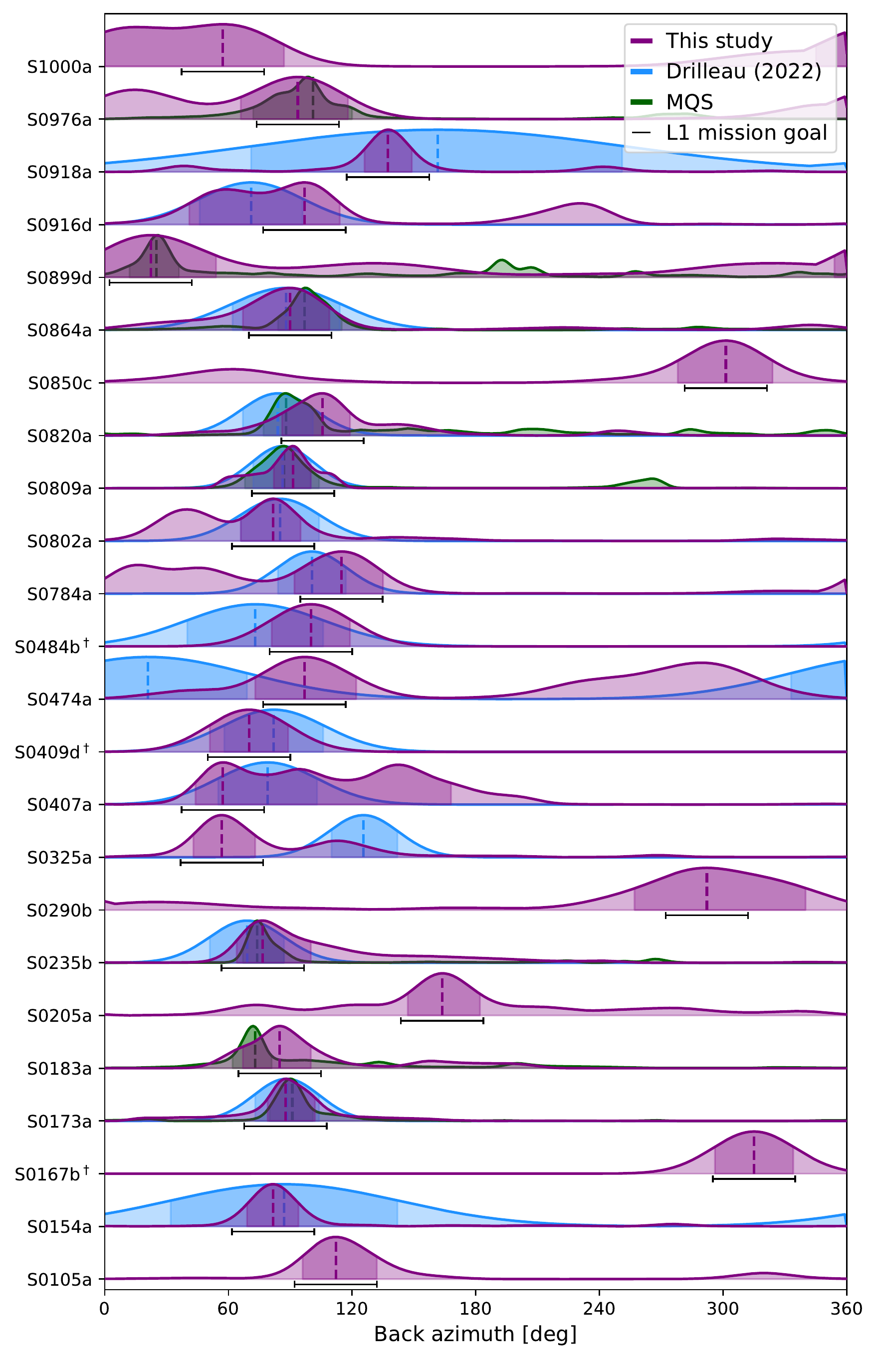}
\caption{Overview of results as seen in Table \ref{Tab:Results_marsquakes}. Probability density functions for this study (purple), \citet{Drilleau2021Locations} (blue), and MQS (green). For this study, we show the KDE for events without marker. Events marked with $\dagger$ have the back azimuth and uncertainty set manually and use a Gaussian curve - for a detailed discussion of each event, see Supplementary Material.}
\label{Fig:results_pdf}
\end{figure}

We also show results from \citet{Drilleau2021Locations}. The authors present their own phase arrival times and do not use the MQS Catalog V8 phase picks, so the results are independent.
We find that our results agree well, with the largest deviation for S0474a (21$^{\circ}$ \citep{Drilleau2021Locations} versus 97$^{\circ}$ (this paper)). For 4 events (S0154a, S0474a, S0484b, S0918a) our method delivers a significantly reduced uncertainty, increasing the value of the result for tectonic interpretation.

\subsection{Comparison with earthquakes and synthetic marsquakes}
\begin{figure}[ht!]
\centering
\noindent\includegraphics[width=0.8\textwidth]{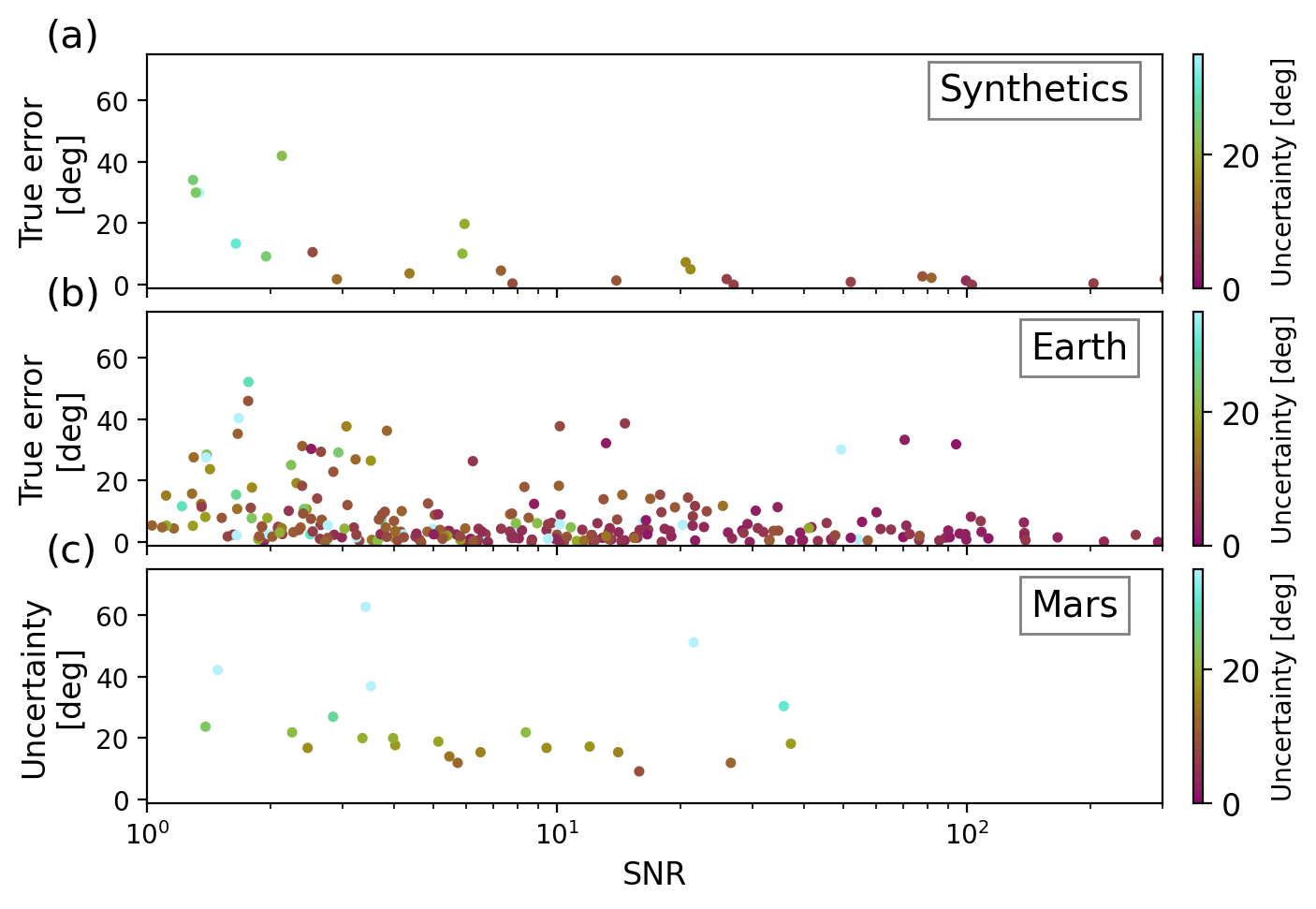}
\caption{Robustness analysis for different types of quakes. (a) True error (difference between estimated back azimuth and true back azimuth) against SNR of the event for synthetic marsquakes. (b) True error against SNR for earthquakes recorded at station WRAB. (c) Estimated error from the width of the KDE peak against SNR for marsquakes. The colors of the dots in the first two rows indicate the estimated error.}
\label{Fig:SNR_vs_error}
\end{figure}

Marsquakes have generally low SNRs compared with earthquakes. To better compare the robustness of the method between different quake sets, we analyse the SNR of each quake and compare it to the error in azimuth estimation. For comparison, we test the terrestrial data set and a synthetic Martian one. We use the earthquake set from station WRAB described previously. Synthetic quakes are generated with magnitudes between $M_w = 2.8-4.2$ and distances between 20$^{\circ}$ and 40$^{\circ}$. This range is representative of intermediate to high quality quakes recorded on Mars. Both sets estimate the back azimuth solely from the P-wave KDE maximum between 0.3--1.0~Hz.
Figure \ref{Fig:SNR_vs_error} shows the error in back azimuth estimation (taken as the absolute difference between true back azimuth and estimated back azimuth) versus SNR for earthquakes and synthetics, and the estimated errors (from the width of the KDE peak) and SNR of marsquakes. Further, we show the estimated error for earthquakes and synthetics for each event. 
The top plot in Figure \ref{Fig:SNR_vs_error} shows the synthetic marsquakes. Most events have errors less than 15$^{\circ}$, with SNR values ranging from around 2 to 80'000 (Figure axis is limited to SNR~=~300). All synthetic quakes with error larger than 20$^{\circ}$ correspond to the lowest magnitude in the data set. Higher SNR values reduce the error in back azimuth estimation. The estimated error decreases with increasing SNR.
The middle plot shows the earthquake set used previously, with 264 events in total. SNR ranges from around 1 to 1000. The largest deviations from the true back azimuth occur when SNR values are low, and decrease when SNR is increasing. Events with low SNR values can still have accurately estimated back azimuths however, with the majority of events having errors below 10$^{\circ}$. The estimated error is comparable to what is seen for the synthetics, though with a larger spread. However, there is no obvious trend visible which would link SNR and estimated error, and the very odd 'blunder' - an event with a large true error with much smaller estimated errors - do occur even for high SNR events.
The bottom plot shows marsquakes, which have generally lower SNR values. Back azimuth and SNR are calculated for an individual frequency range for each event depending on frequency content and glitch presence, making direct comparisons not straightforward in terms of SNR. The estimated errors are similar to what is estimated for the earthquake set, barring some outliers there. The marsquake set in general is comparable to low-SNR earthquakes in terms of estimated error.

\subsection{HF and VF marsquakes}
In contrast to the previously discussed LF and BB marsquakes, the HF and VF events have their main energy content above 1~Hz, sometimes reaching 30~Hz and higher. Since these high frequencies are dominated by scattering, and several prominent lander modes are present between 1 and 9~Hz, it is much more challenging to find a stable polarization. We try the same methodology but with an adapted frequency band of 0.25--1.25~Hz for the KDE. This broader band, which extends to higher frequencies, should better capture the characteristics of these events. While their main energy is at high frequencies, for some larger events there is a tail of energy below 1~Hz. The amplitude of this differs between events. At frequencies much above our selected band, scattering is dominant and a stable polarization is not expected. Unfortunately, we do not find a stable back azimuth for any high frequency family event recorded so far. This suggests that the arrival of a ballistic body wave is not observed above noise level and that the first arrival is already strongly scattered \citep{vanDriel2021_HF, menina_energy_2021}.

\section{Discussion}
Polarization analysis has been used to improve seismic signal detection and to study Earth's microseism. As shown in \citet{stutzmann2021polarized}, the ambient noise on Mars is mainly linear in polarization, due to local wind noise. However,  polarization analysis provides a suitable tool to assess weak or complex seismic signals, such as marsquakes. Only few back azimuths of marsquakes have been determined so far by MQS, severely limiting tectonic interpretation from the MQS catalog. More complex techniques are needed to obtain back azimuth estimates for previously unlocated events, and to improve back azimuth estimates of already located events.
The polarization analysis presented in this paper is capable of reliably estimating the back azimuth of seismic events with sufficient quality from the P and S-wave arrivals. 
Analysis of synthetic marsquakes shows that the back azimuth error is generally below 10$^{\circ}$ for quakes with distances and magnitudes similar as those recorded on Mars. For small SNRs ($<2$), the error can reach around 40$^{\circ}$. While the background noise recorded on Mars contains several features which interfere with the analysis of seismic data, we find that the removal of the 1~Hz tick noise is sufficient to retrieve the true back azimuth in the desired frequency range.
The earthquake set recorded at station WRAB in Tennant Creek, Australia, shows a similar pattern. Over 260 earthquakes are analysed, with SNRs ranging from 1 to 1000. Similar to the synthetic quakes, we see an improved estimated back azimuth for events with higher SNR. The majority of events have an error of less than 10$^{\circ}$, with maximum errors just over 40$^{\circ}$. The estimated uncertainty from the polarization analysis is larger for the earthquake data set than for the synthetic set. While the uncertainty covers the true back azimuth in the majority of cases, it is sometimes estimated too narrow.

Marsquake data has its own characteristics (e.g. tick noise, glitches, resonances and lander modes, amongst others) and thus requires careful handling to extract the desired signal.
We applied the same method to a set of high quality marsquakes to estimate their back azimuths. The polarization back azimuths are close to the back azimuths estimated by MQS and \citet{Drilleau2021Locations} for almost all events. We also provide back azimuth estimates for a number of events where no MQS back azimuth is available. Most of these are more speculative however, owing to poorer event quality and/or low SNR. Some events are contaminated by glitches close to the phase arrivals. However, careful analysis of both P and S-wave polarization and considering the perpendicularity constraint, a back azimuth can still be estimated, albeit with less certainty. We place the majority of these previously unlocated events in a similar, eastern, direction as most of the better constrained events. We combine our results with distances provided by MQS \citep{MQSCatalog} to obtain locations for the events (Fig. \ref{Fig:Map}). Within uncertainties, we therefore confirm that most seismic activity near InSight happens in the Cerberus Fossae graben system \citep{banerdt_initial_2020}. However, we also identify new active regions. S0899d heralds a source region due North, in Utopia Planitia; S0290b and S0850c, lie north-west, near Isidis basin. S0205a and S0918a are the only events that seem to originate from southern hemisphere, though S0205a in particular is a challenging event to locate. Neither S0205a nor S0899d has a second phase pick, making their distances uncertain. As both events are similar to S0183a, we place them at similar distances (46\textdegree) \citep{CeylanV9Catalog}.
The two most distant events recorded by InSight, S0976a and S1000a, are located in Valles Marineris and the northern Tharsis region (albeit with a large uncertainty), respectively. S0167b could be located close to the boundary between Utopia Planitia and the polar Vastitas Borealis.
While the parameters used in this paper seem well suited to estimate event back azimuths, future marsquakes may need adjustments based on event characteristics.

\section{Conclusion}
We showed that back azimuth determination can be improved by a combination of eigenvector analysis on both P and S-arrivals. This study increases the amount of information on source regions that is possible with the InSight data set. The method allows for an automated determination and quality control of back azimuths from single stations that could also be applied to remote and sparse networks on Earth.

We found that the majority of marsquakes located with this method are originating from the general Cerberus Fossae region. A few events seem to originate from different directions. The two most distant marsquakes are located in the Tharsis region on Mars, with one quake possibly in Valles Marineris. The new locations will enrich tectonic interpretation of the single-plate planet Mars. InSight continues to record the Martian seismicity and expand the Martian seismic catalog. Each additional event with a constrained location will help build a more comprehensive picture of Mars' tectonic activity and its interior.

\section*{Acknowledgements}
We acknowledge funding from ETH Zurich through the ETH+ funding scheme (ETH+02 19-1: ``Planet Mars"). We acknowledge NASA, CNES, partner agencies and institutions (UKSA, SSO, DLR, JPL, IPGP-CNRS, ETHZ, ICL, MPS-MPG), and the operators of JPL, SISMOC, MSDS, IRIS-DMC and PDS for providing SEED SEIS data.
This is InSight contribution 221.

\section*{Data and Resources}
The InSight seismic event catalog Version 9 \citep{MQSCatalog}, waveform data and station metadata are available from IRIS-DMC and the IPGP datacenter \citep{network_XB}. 
Earthquake data were obtained from seismic network II \citep{network_II} and accessed with ObsPyDMT \citep{obspyDMT} using the IRIS catalog.
Synthetic seismograms were calculated using the Instaseis \citep{Instaseis2015} Martian model \textit{InSight\_KKS21GP} \citep{KKS_2021GP} which can be accessed at \url{http://instaseis.ethz.ch/marssynthetics/InSight_KKS21_GP/}. Seismic data was handled with ObsPy \citep{beyreuther2010obspy, Krischer_2015}. Calculations in Python were done with Numpy \citep{Numpy} and Scipy \citep{Scipy} and the results were visualised with Matplotlib \citep{Matplotlib} and seaborn \citep{seaborn}.
The Supplementary Material contains a detailed polarization analysis for each marsquake in Table \ref{Tab:Results_marsquakes} not presented in the main text.

\section*{Declaration of Competing Interests}
The authors acknowledge there are no conflicts of interest recorded.

\bibliographystyle{apalike}
\bibliography{references.bib}

\clearpage
\appendix
\setcounter{page}{1}
\renewcommand{\thefigure}{S\arabic{figure}}
\setcounter{figure}{0} 
\section*{Supplementary Material}

{\Large Low Frequency Marsquakes and Where to Find Them: Back Azimuth Determination Using a Polarization Analysis Approach}\\

G\'{e}raldine Zenh\"ausern, Simon C. St\"ahler, John F. Clinton, Domenico Giardini, Savas Ceylan, Rapha\"{e}l F. Garcia

\begin{figure}[ht]
\centering
\noindent\includegraphics[width=\textwidth]{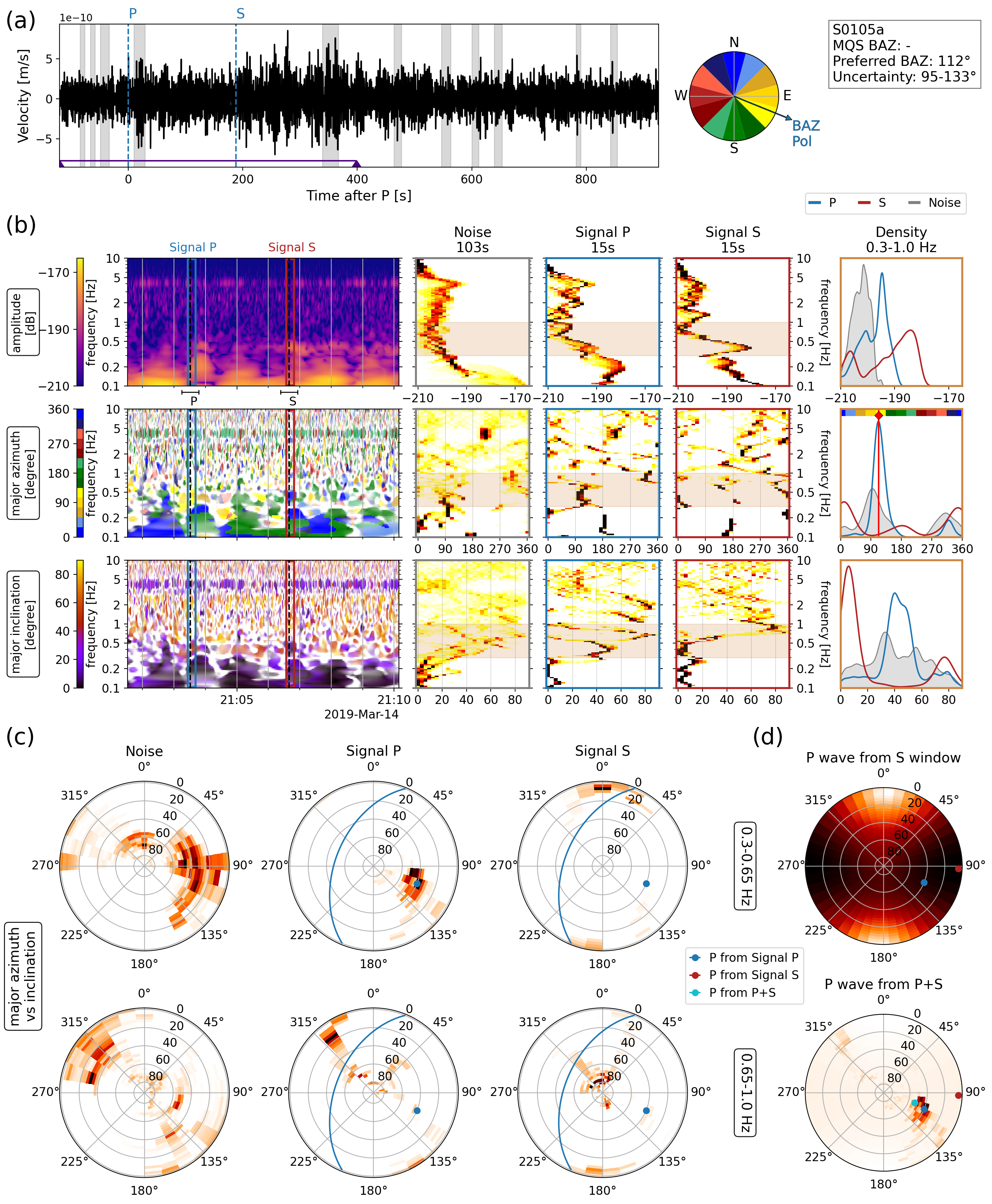}
\caption{Polarization analysis for marsquake S0105a ($M_w$3.0 LF QB event at 32.5$^{\circ}$ distance, March 14, 2019). The plot follows the same structure as Fig. \ref{Fig:Syn_polarization}.
The frequency band of 0.3--0.65~Hz shows moderate to high inclination values for the P window, as well as low inclination values for the S window. At frequencies below 0.3 Hz, P is dominated by a glitch 30 seconds after the picked arrival. The S-arrival has a strong polarization at 0\textdegree and 190\textdegree, which is consistent with the P-polarization, but suggests that the value could be 10-20\textdegree too high.}
\label{Fig:Appendix_S0105a}
\end{figure}

\begin{figure}[ht]
\centering
\noindent\includegraphics[width=\textwidth]{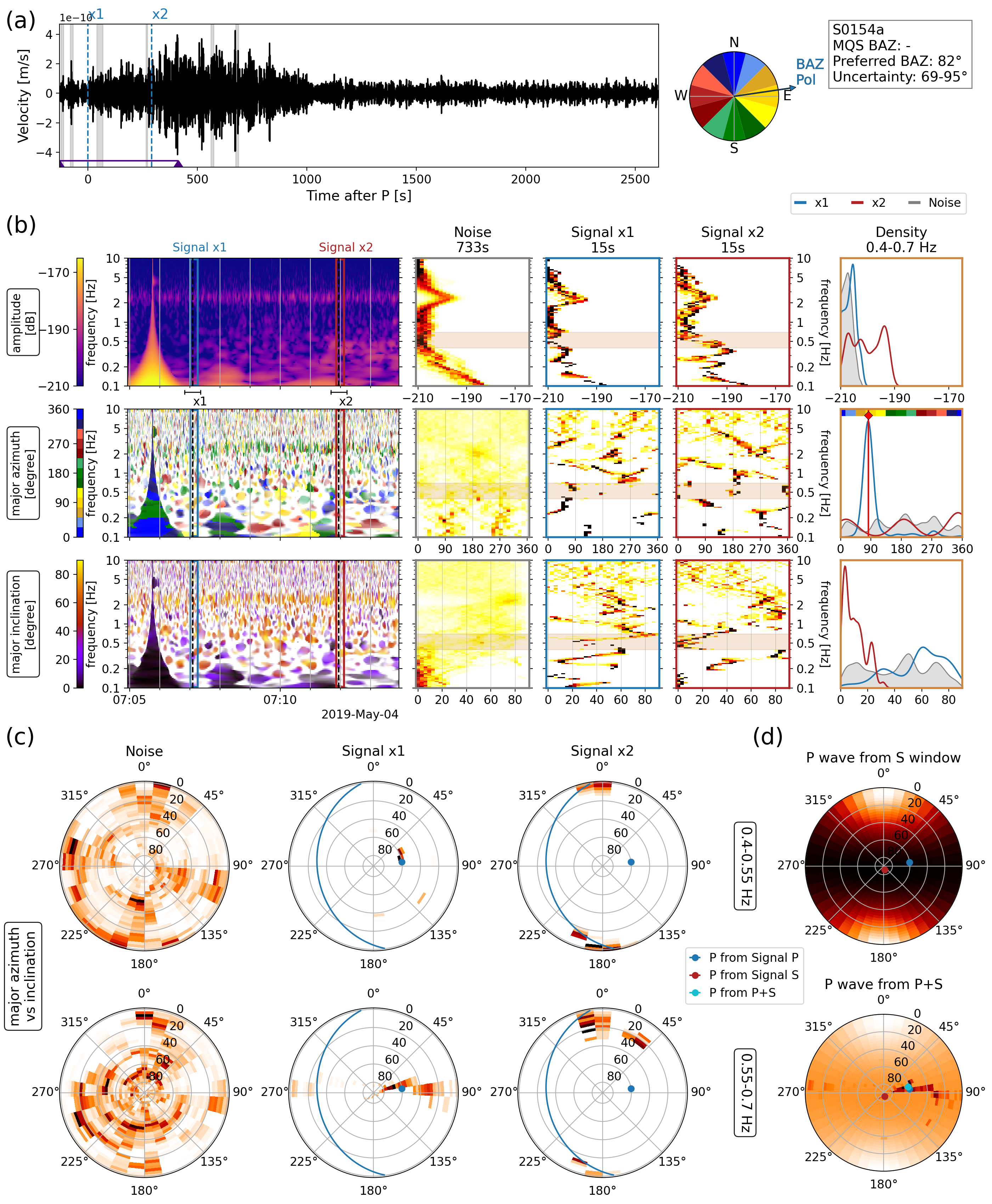}
\caption{Polarization analysis for marsquake S0154a ($M_w$2.9 BB QC event at 33.7$^{\circ}$ distance, May 4, 2019). The plot follows the same structure as Fig. \ref{Fig:Syn_polarization}. The two arrivals are marked as x1 and x2 by MQS since they are not unequivocally P and S using standard tools. In our analysis, we find that they can be identified well as P and S in a relatively narrow frequency band from 0.4--0.7~Hz. x2 has a high-inclination signal at 0.3--0.4~Hz, which seems inconsistent with S, but could be explained by the low SNR. Its horizontal signal is consistent with a back azimuth from the east (d). A weaker x1 and some contamination of x2 make this event challenging, though results are consistent for both phases.}
\end{figure}

\begin{figure}[ht]
\centering
\noindent\includegraphics[width=\textwidth]{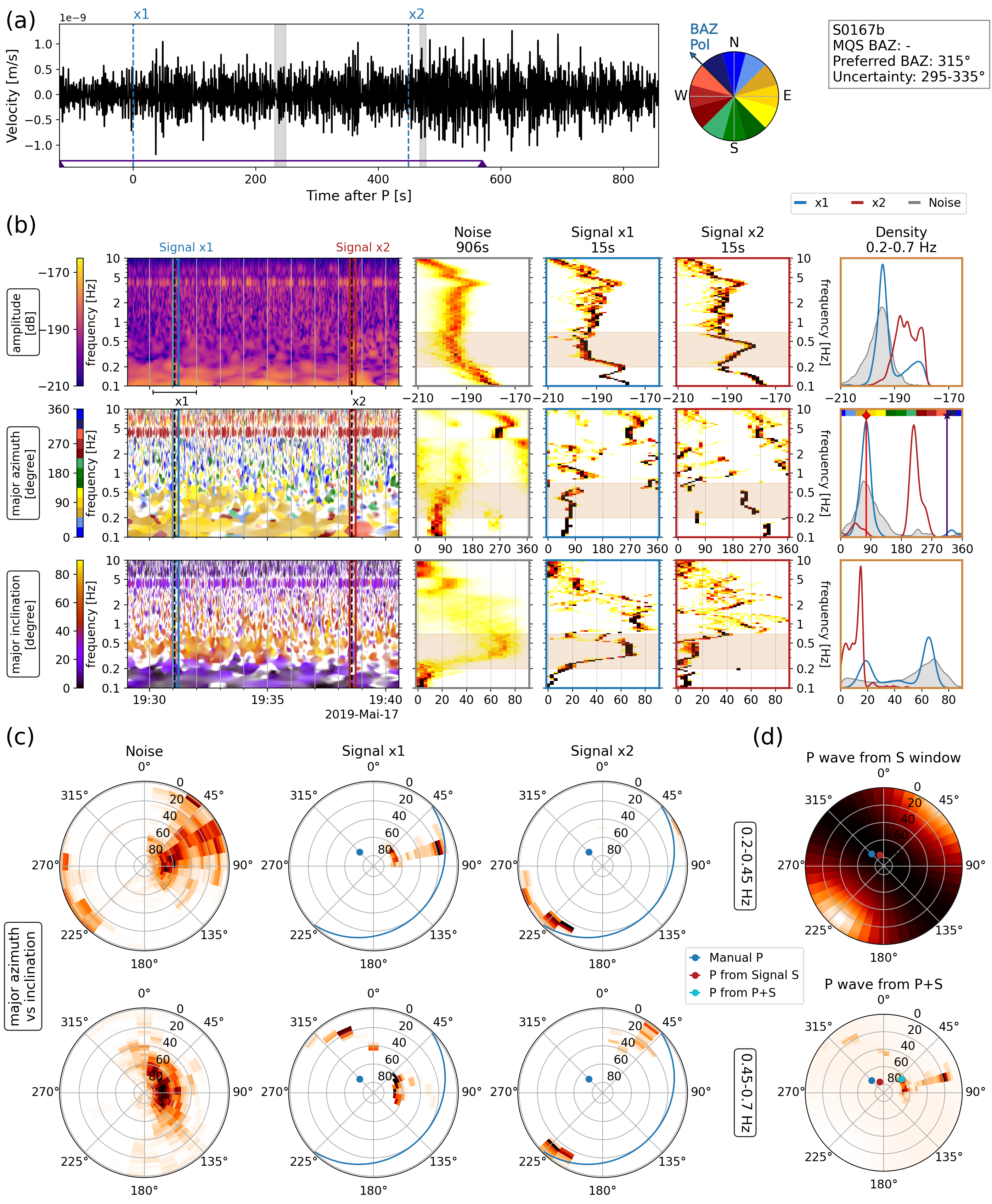}
\caption{Polarization analysis for marsquake S0167b ($M_w$ unknown LF QC event at unknown distance, May 17, 2019). The plot follows the same structure as Fig. \ref{Fig:Syn_polarization}. This event is at a larger distance than most others discussed in this article. The identification of x1 as P is difficult based on polarization only, since the direction does not differ significantly from the pre-event noise window, which is contaminated by wind. x2 shows a low-inclination signal at azimuth 40 or 230\textdegree, distinct from the wind-induced noise before. Its signal is also significantly stronger than the noise. We assign a manual back azimuth of 315\textdegree, consistent with the polarization of 300-340\textdegree from x2.}
\end{figure}

\begin{figure}[ht]
\centering
\noindent\includegraphics[width=\textwidth]{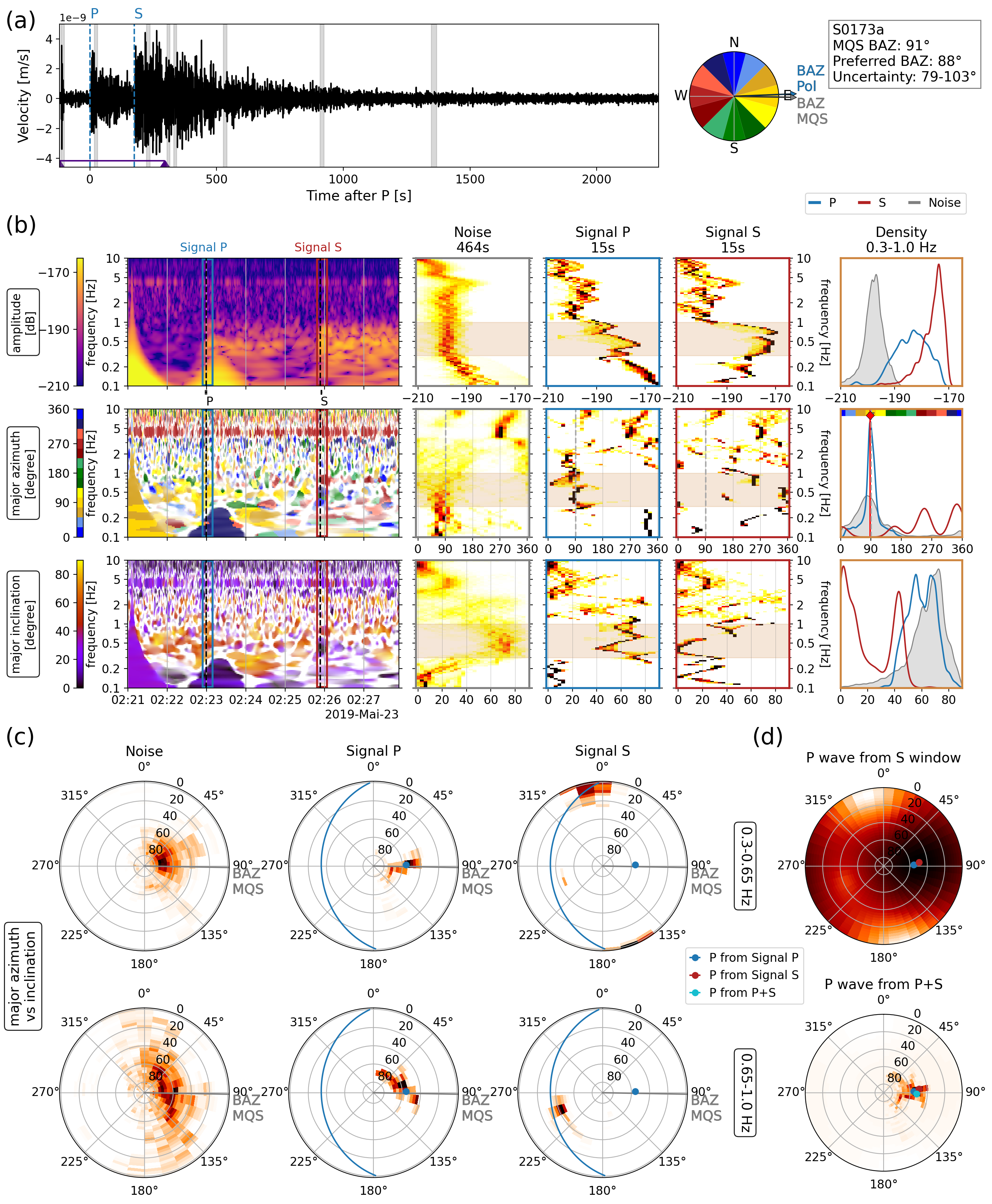}
\caption{Polarization analysis for marsquake S0173a ($M_w$3.7 LF QA event at 30$^{\circ}$ distance, May 23, 2019). The plot follows the same structure as Fig. \ref{Fig:Syn_polarization}. One of the best events recorded by InSight during the mission so far. The P is contaminated by a glitch arriving 20 seconds after, visible by a strongly horizontal signal (blue in the azimuth plot). Using frequencies between 0.3--1.0~Hz avoids the glitch. Both P and S are strongly polarized and show consistent results of 90\textdegree.}
\end{figure}

\begin{figure}[ht]
\centering
\noindent\includegraphics[width=\textwidth]{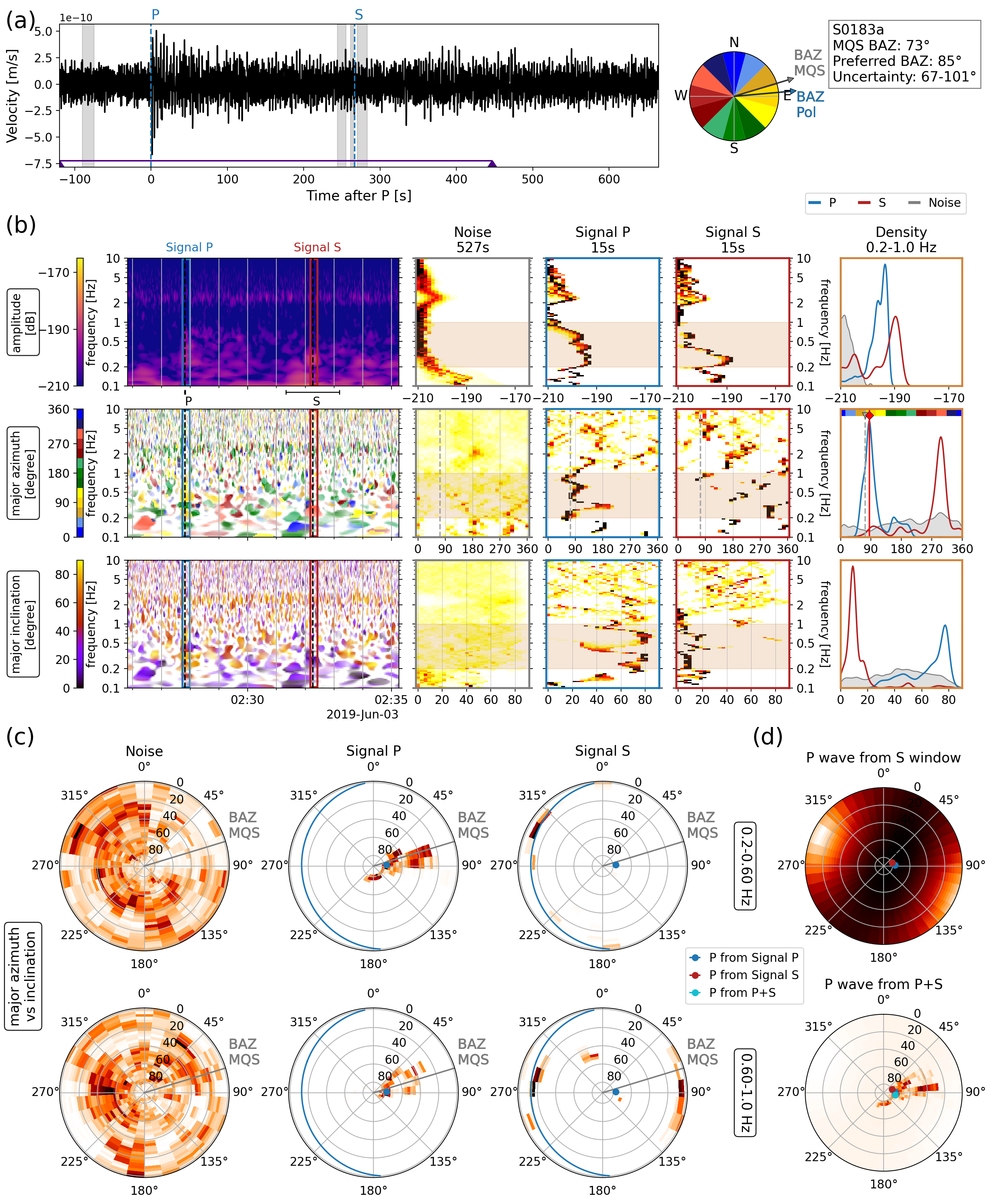}
\caption{Polarization analysis for marsquake S0183a ($M_w$3.1 LF QB event at 46.4$^{\circ}$ distance, June 3, 2019). The plot follows the same structure as Fig. \ref{Fig:Syn_polarization}. The P-wave has a consistent polarization between 0.2--1.0~Hz, with a strongly vertical signal. The S-wave has a horizontal inclination in the same frequency band, though it is possibly glitch-contaminated (see (a) and \citet{clinton2021_catalog} for a detailed discussion). Results from P and the presumptive S are consistent with each other.}
\end{figure}

\begin{figure}[ht]
\centering
\noindent\includegraphics[width=\textwidth]{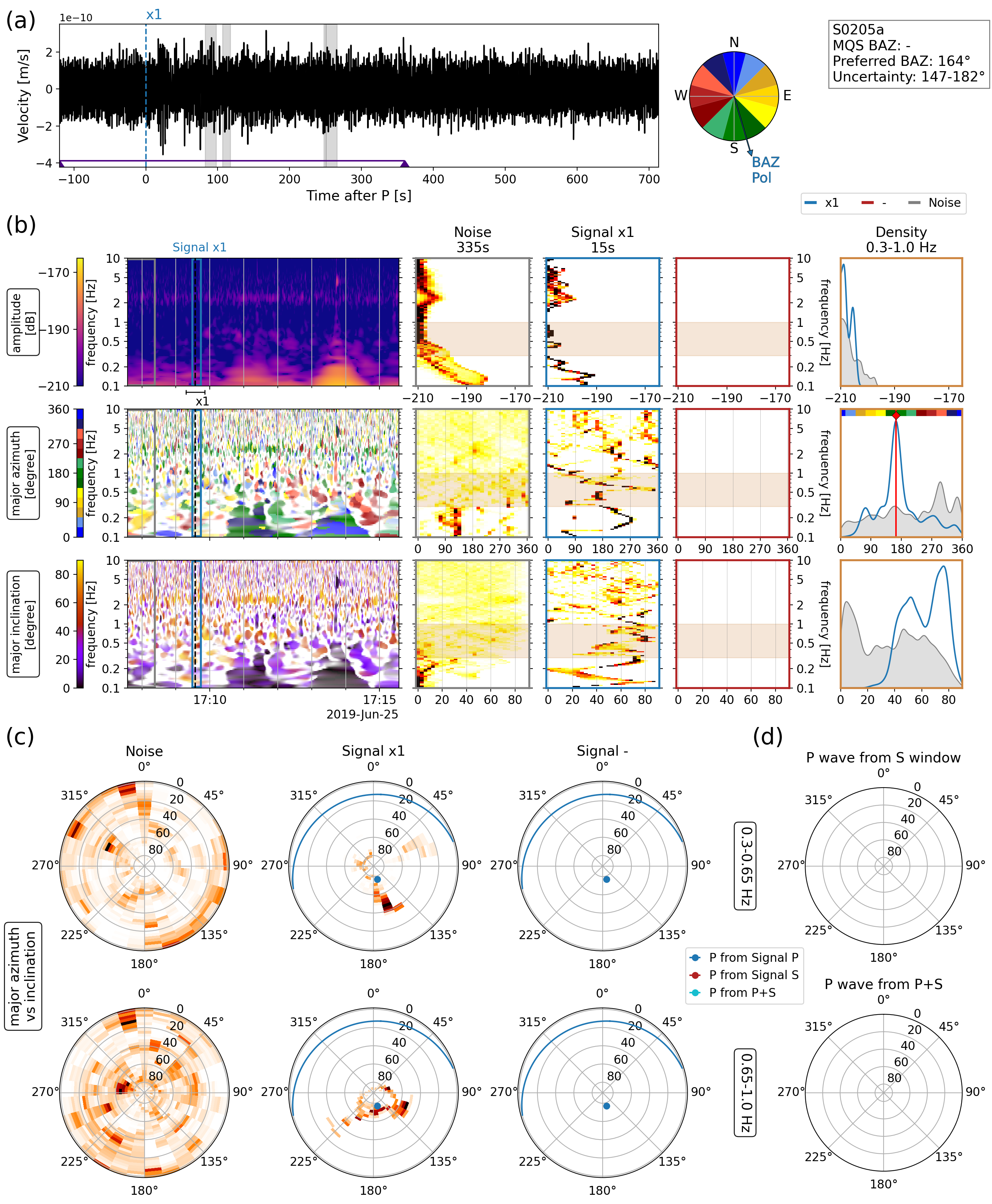}
\caption{Polarization analysis for marsquake S0205a ($M_w$3.0 BB QC event at 42.4$^{\circ}$ distance, June 25, 2019). The plot follows the same structure as Fig. \ref{Fig:Syn_polarization}. The event lacks a second pick, so only x1 is available. x1 shows a high inclination between 0.3--1.0~Hz, and a southward polarization. Since no second pick is available, the distance is obtained from alignment \citep{MQSCatalog}.}
\end{figure}

\begin{figure}[ht]
\centering
\noindent\includegraphics[width=\textwidth]{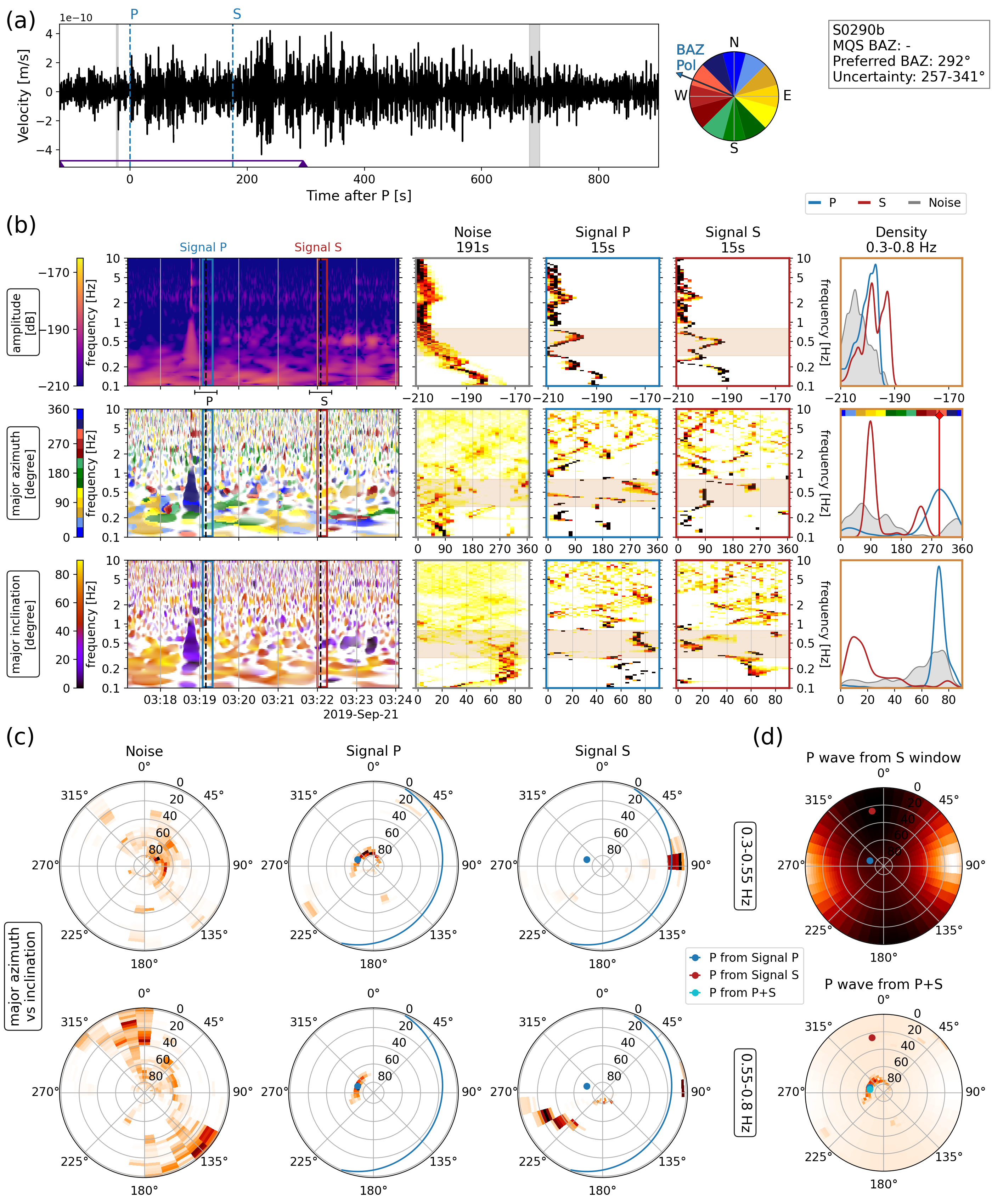}
\caption{Polarization analysis for marsquake S0290b ($M_w$3.5 LF QB event at 30.4$^{\circ}$ distance, September 21, 2019). The plot follows the same structure as Fig. \ref{Fig:Syn_polarization}. There is a strong glitch shortly before the P with horizontal inclination, but the P window is not affected (cf. inclination row). The S window contains a clear horizontal signal between 0.4 and 0.7 Hz, clearly different from the largely vertical energy present otherwise. The frequency band 0.3--0.8~Hz catches this horizontal energy while containing the stable azimuth signal in the P window. There is general agreement between the P and S results as seen in (d), resulting in an estimated back azimuth of 257-341 \textdegree.}
\end{figure}

\begin{figure}[ht]
\centering
\noindent\includegraphics[width=\textwidth]{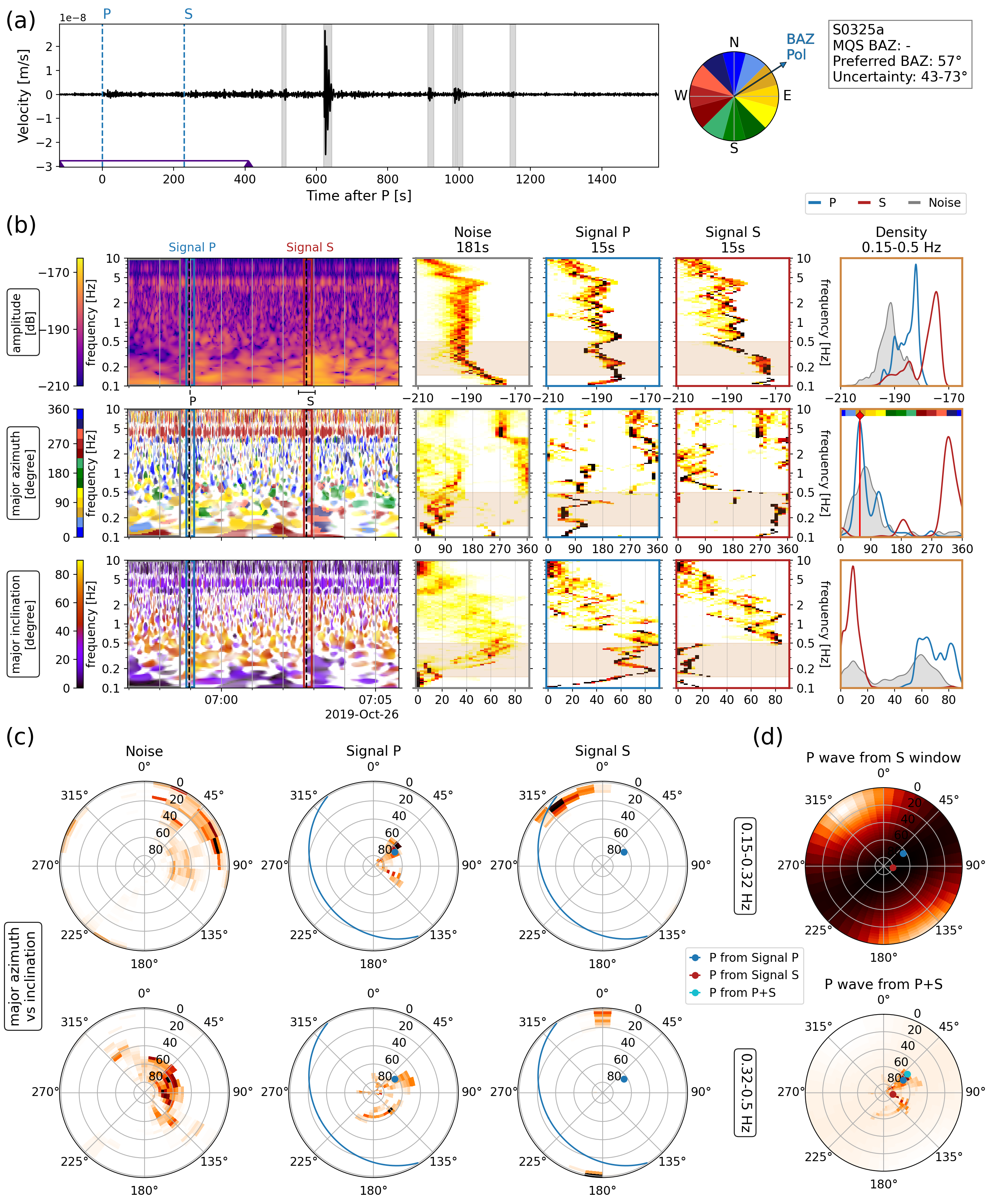}
\caption{Polarization analysis for marsquake S0325a ($M_w$3.7 LF QB event at 39.7$^{\circ}$ distance, October 26, 2019). The plot follows the same structure as Fig. \ref{Fig:Syn_polarization}. The frequency band of 0.15--0.5~Hz contains both the high inclination part of the P window and the low inclination part of the S window. As seen in (c), the P-wave window has a polarization that is significantly different from the noise window below 0.3 Hz. While the noise above 0.3 Hz has a similar azimuth as the P window, agreement with the S window and higher inclination suggest that the P-wave polarization is robust and shows the true event back azimuth of 43-73\textdegree.}
\end{figure}

\begin{figure}[ht]
\centering
\noindent\includegraphics[width=\textwidth]{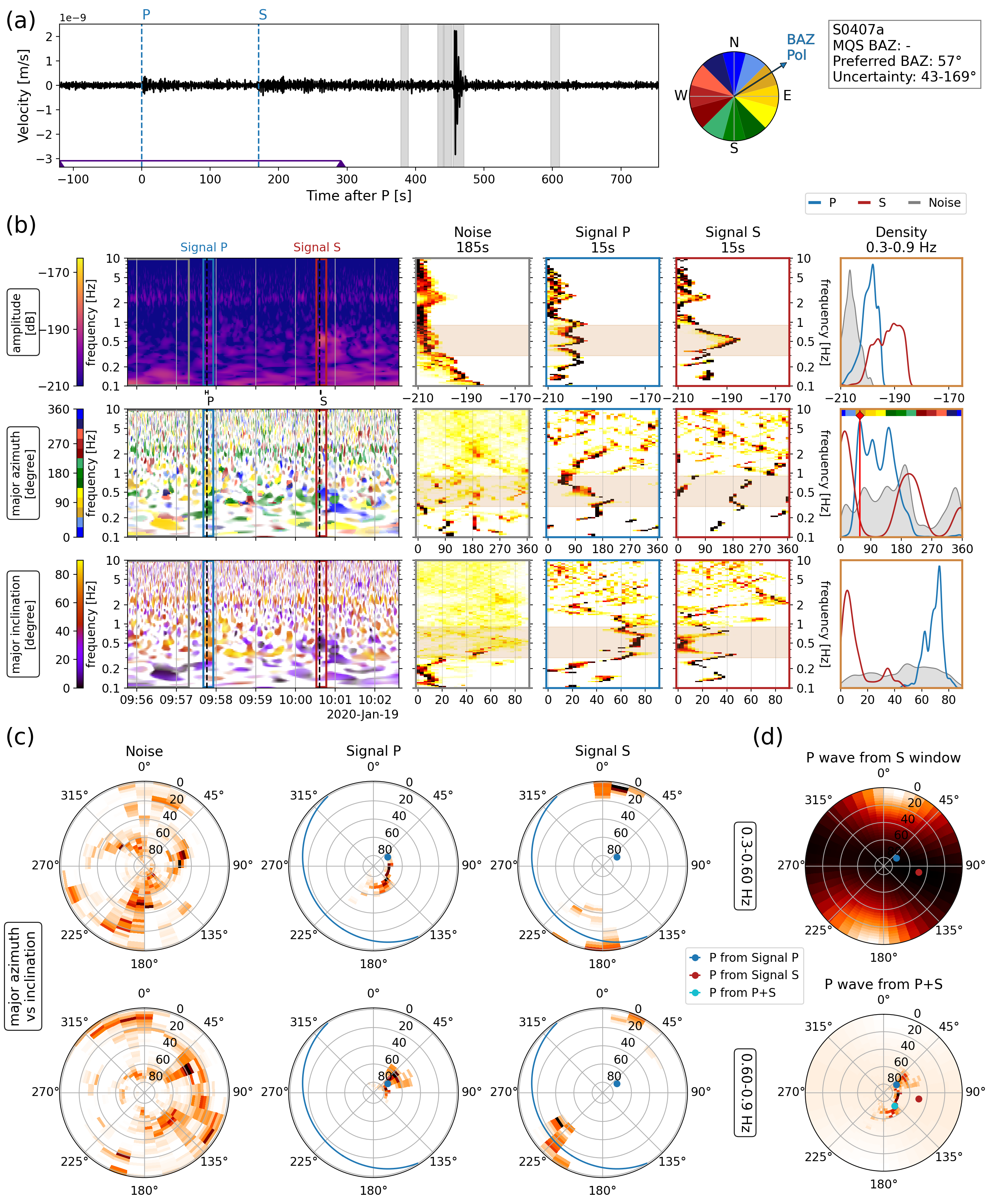}
\caption{Polarization analysis for marsquake S0407a ($M_w$2.9 LF QB event at 29.3$^{\circ}$ distance, January 19, 2020). The plot follows the same structure as Fig. \ref{Fig:Syn_polarization}. There is clear vertical and horizontal energy visible at the P and S arrival, respectively. The P-wave has a narrow-peaked azimuth distribution above 0.6 Hz, but is more broad below. The S-wave polarization shows the opposite picture: Peaks at 0 and 180\textdegree below 0.6 Hz (consistent with an event BAZ of 90\textdegree), but 30\textdegree and 210\textdegree above 0.6 Hz (consistent with BAZ of 100-120\textdegree. In summary, the event is likely located due east, but with a relatively high uncertainty, if one takes all frequencies into account.}
\end{figure}

\begin{figure}[ht]
\centering
\noindent\includegraphics[width=\textwidth]{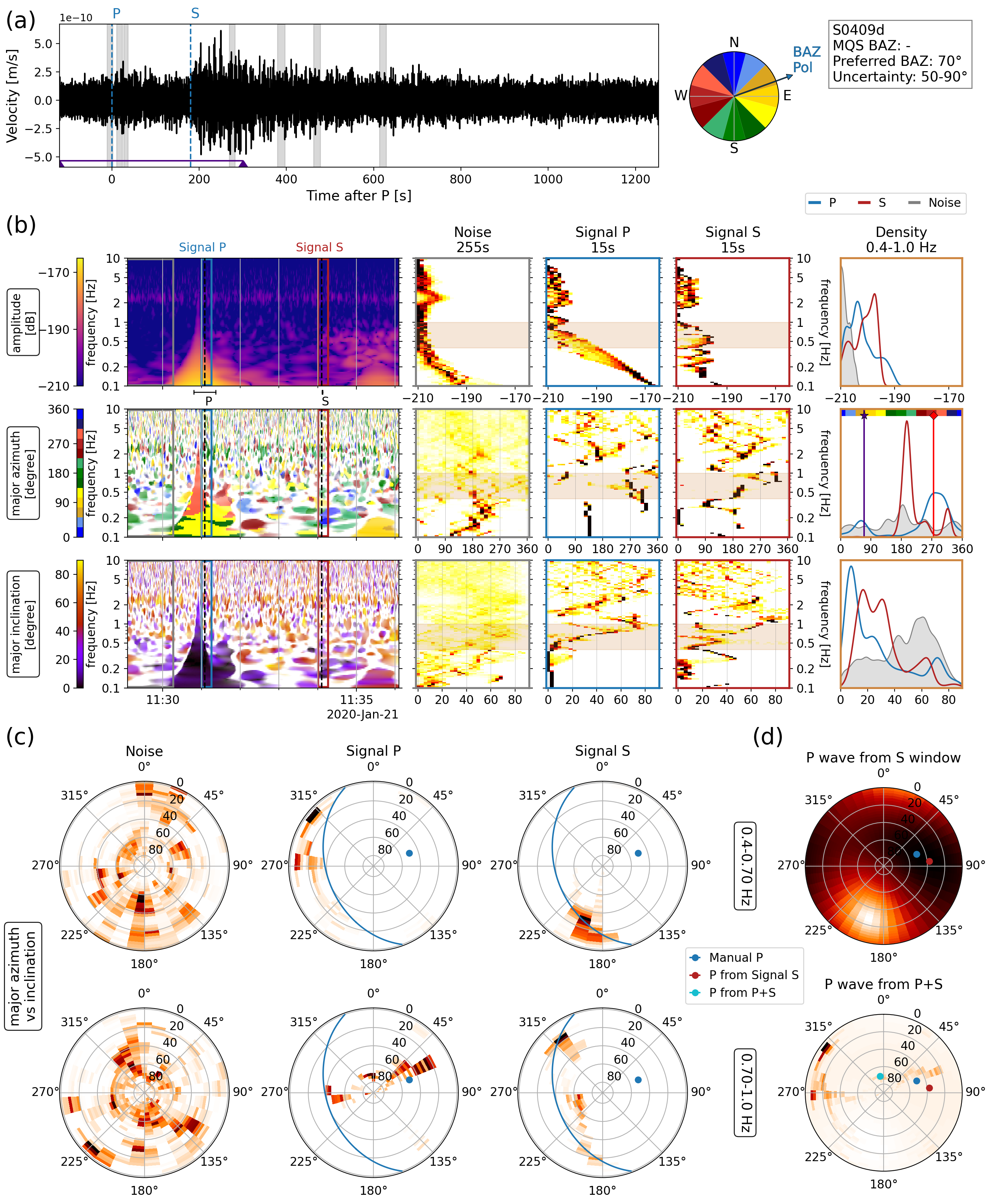}
\caption{Polarization analysis for marsquake S0409d ($M_w$3.2 LF QB event at 31.1$^{\circ}$ distance, January 21, 2020). The plot follows the same structure as Fig. \ref{Fig:Syn_polarization}. The P arrival is dominated by several large glitches, and the event signal is difficult to retrieve. The S arrival shows clear horizontal energy and it suggests a back azimuth around 80$^{\circ}$. Analysing the P window more closely with this knowledge shows vertical energy in a similar direction. The back azimuth is manually set to 70$^{\circ}$, and (c) and (d) show that there is good agreement with the S window for this.}
\end{figure}

\begin{figure}[ht]
\centering
\noindent\includegraphics[width=\textwidth]{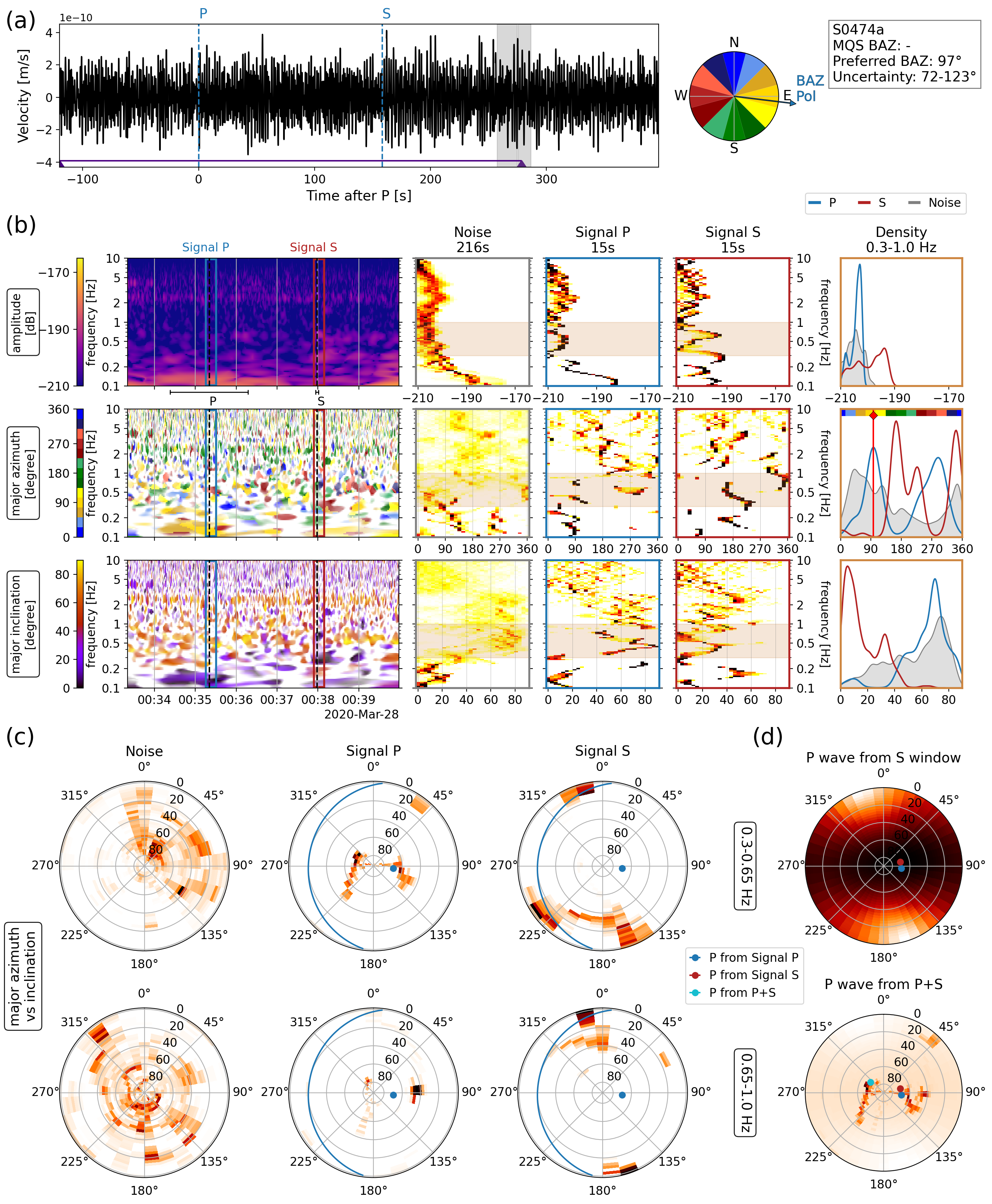}
\caption{Polarization analysis for marsquake S0474a ($M_w$2.9 LF QC event at 29.1$^{\circ}$ distance, March 28, 2020). The plot follows the same structure as Fig. \ref{Fig:Syn_polarization}. The P-wave pick is speculative with $\pm$60~s uncertainty. The amplitude KDE shows that the P window has no strong amplitude, resulting in a very weak SNR based on the P. However, the S window shows both a stronger amplitude and clear horizontal energy, suggesting a back azimuth of around 90$^{\circ}$.}
\end{figure}

\begin{figure}[ht]
\centering
\noindent\includegraphics[width=\textwidth]{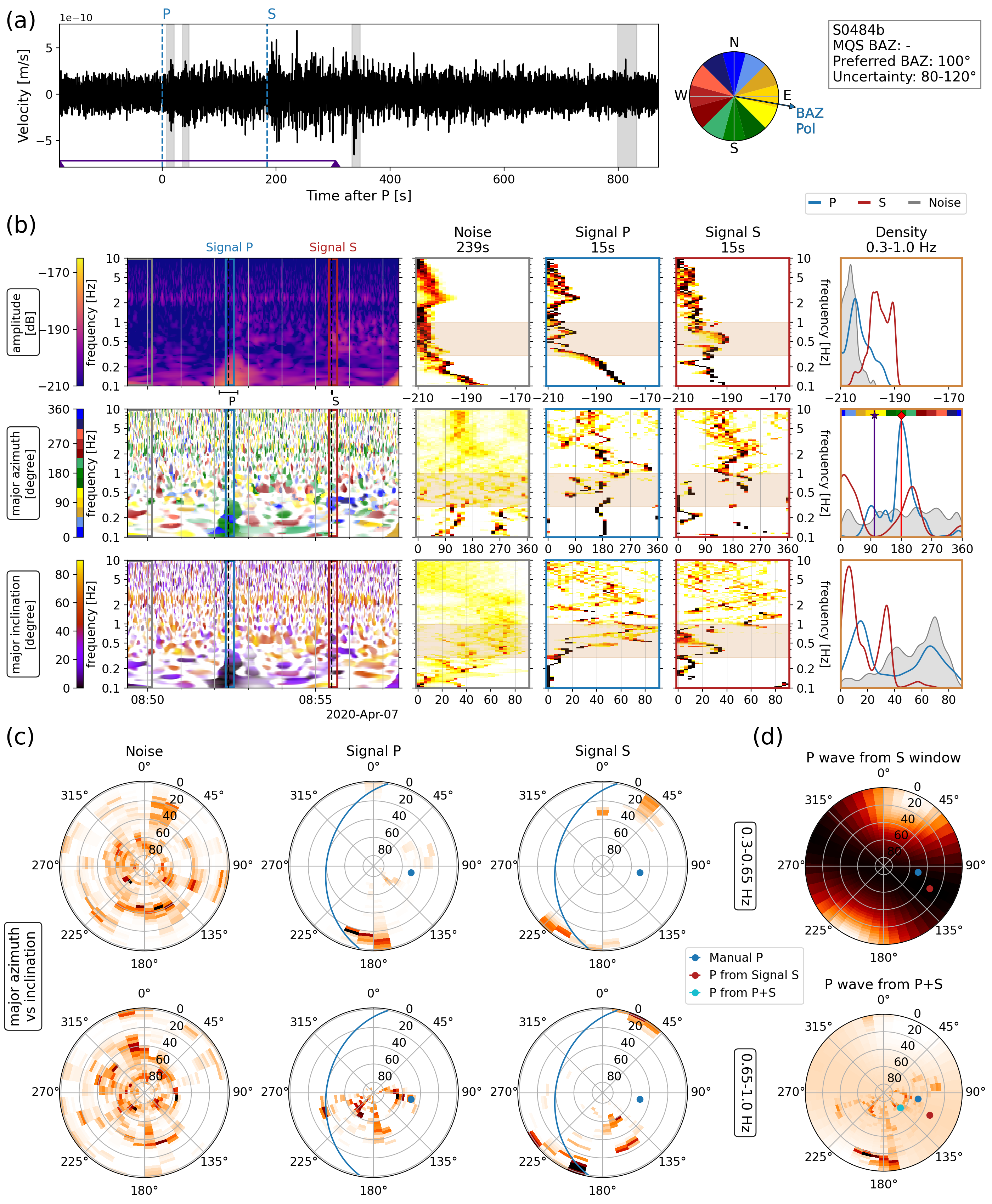}
\caption{Polarization analysis for marsquake S0484b ($M_w$2.9 BB QB event at 31.8$^{\circ}$ distance, April 7, 2020). The plot follows the same structure as Fig. \ref{Fig:Syn_polarization}. The P-arrival is contaminated by a strong glitch (visible in (a) and (b)). There is some vertical energy visible in the P window at higher frequencies. The S window has a clear horizontal arrival with strong amplitude. It suggests a back azimuth around 110$^{\circ}$. The back azimuth is set to 100$^{\circ}$ after careful analysis of the P window.}
\end{figure}

\begin{figure}[ht]
\centering
\noindent\includegraphics[width=\textwidth]{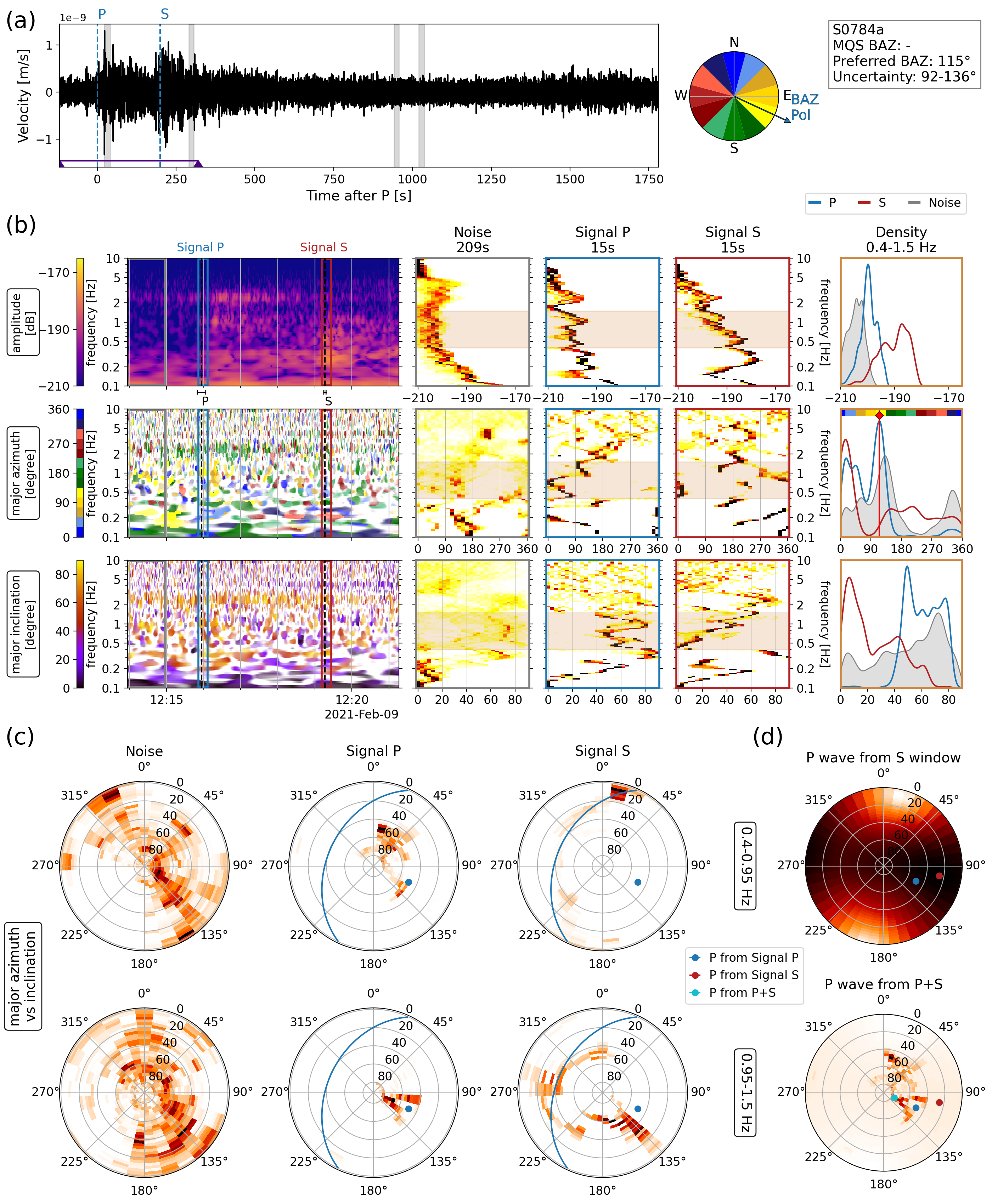}
\caption{Polarization analysis for marsquake S0784a ($M_w$3.3 BB QB event at 34.5$^{\circ}$ distance, February 9, 2021). The plot follows the same structure as Fig. \ref{Fig:Syn_polarization}. Both P and S arrival show clear vertical and horizontal energy respectively. The S window shows the horizontal energy as more dominant at lower frequencies, while the P window shows dominantly vertical energy over the whole frequency band. Results from both windows are consistent around 115$^{\circ}$.}
\end{figure}

\begin{figure}[ht]
\centering
\noindent\includegraphics[width=\textwidth]{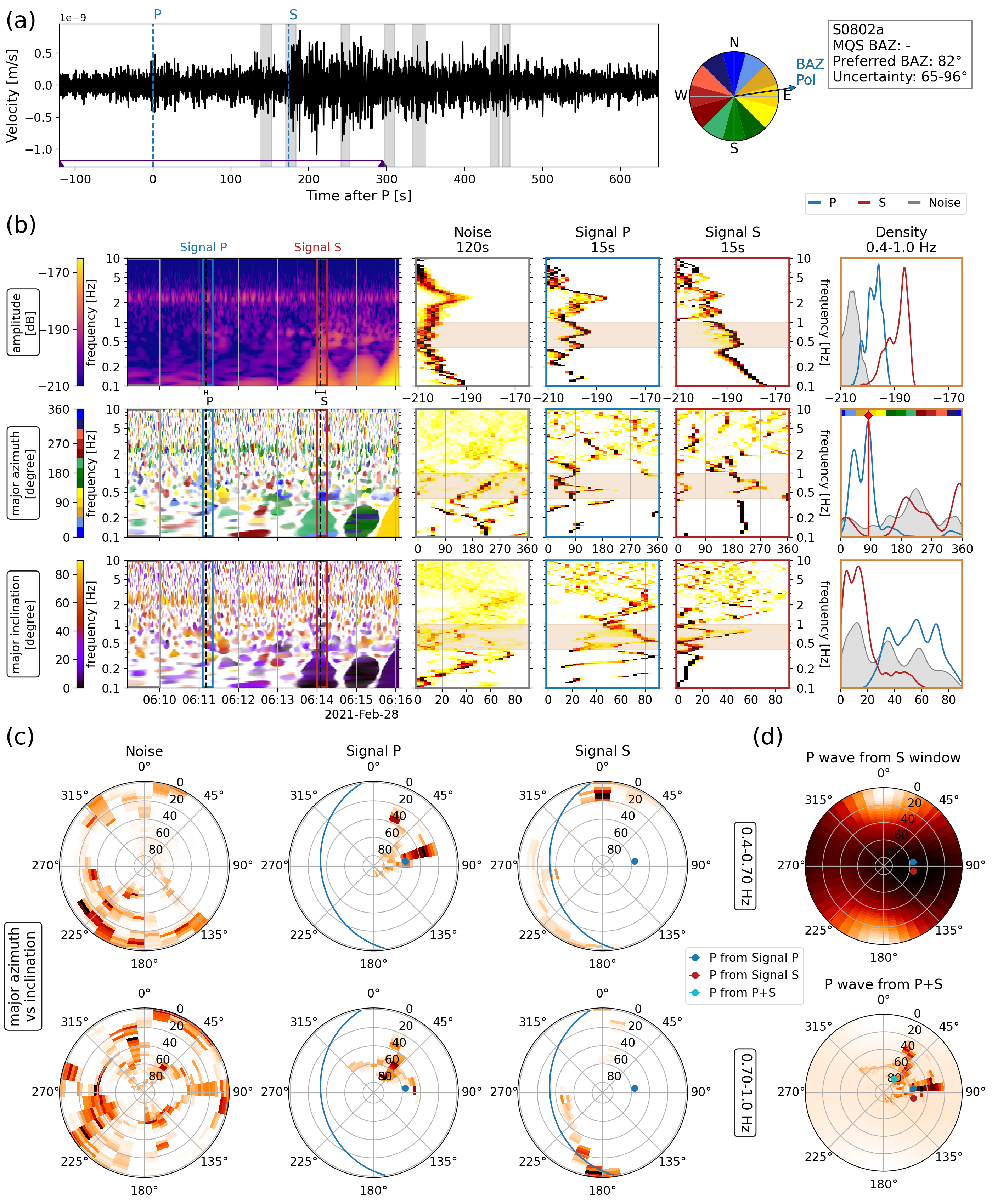}
\caption{Polarization analysis for marsquake S0802a ($M_w$2.9 BB QB event at 30$^{\circ}$ distance, February 28, 2021). The plot follows the same structure as Fig. \ref{Fig:Syn_polarization}. The P window shows a strongly polarized signal, predominately vertical, at 80\textdegree below 0.7 Hz and at 40\textdegree above, distinct from the pre-event noise. The S-wave arrival is contaminated by a glitch, though the frequency band of 0.4--1.0~Hz seems to avoid the bulk of it (as evident from the more scattered inclination and the differing azimuth). Both signal windows have amplitudes clearly above the noise, and the azimuth and inclination are different as well. Results from both windows are consistent with 82\textdegree BAZ}
\end{figure}

\begin{figure}[ht]
\centering
\noindent\includegraphics[width=\textwidth]{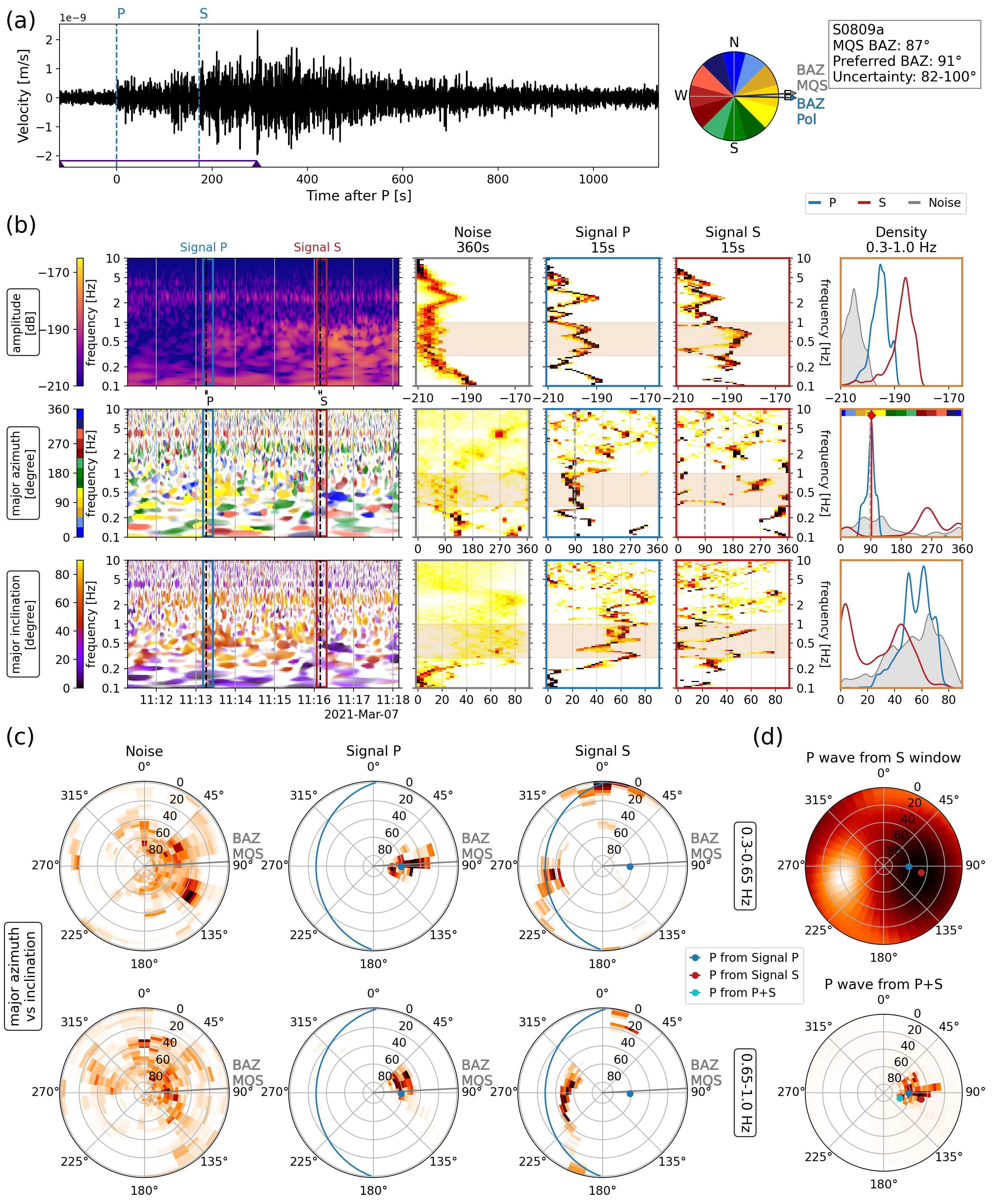}
\caption{Polarization analysis for marsquake S0809a ($M_w$3.3 LF QA event at 29.8$^{\circ}$ distance, March 7, 2021). The plot follows the same structure as Fig. \ref{Fig:Syn_polarization}. Both P and S arrivals are clear with no glitch contamination. The P-wave polarization is similar to the pre-event noise, but the S-wave polarization is distinct. They both show a consistent polarization which agree with each other.}
\end{figure}

\begin{figure}[ht]
\centering
\noindent\includegraphics[width=\textwidth]{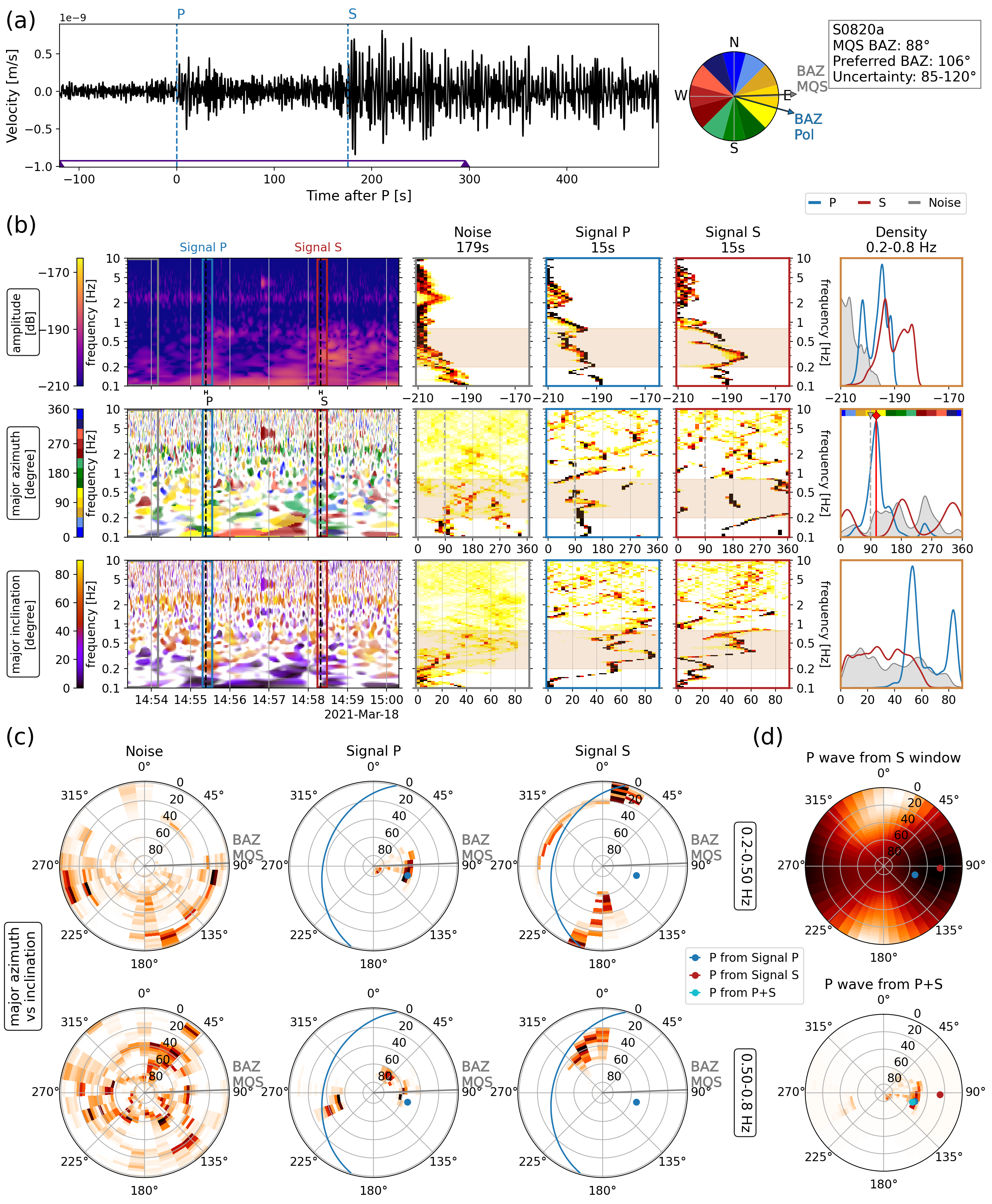}
\caption{Polarization analysis for marsquake S0820a ($M_w$3.3 LF QA event at 30.2$^{\circ}$ distance, March 18, 2021). The plot follows the same structure as Fig. \ref{Fig:Syn_polarization}. Both P and S arrivals are clear with no glitch contamination. The P-wave polarization is similar to the pre-event noise, but the S-wave polarization is distinct. They both show a consistent polarization which agree with each other.}
\end{figure}

\begin{figure}[ht]
\centering
\noindent\includegraphics[width=\textwidth]{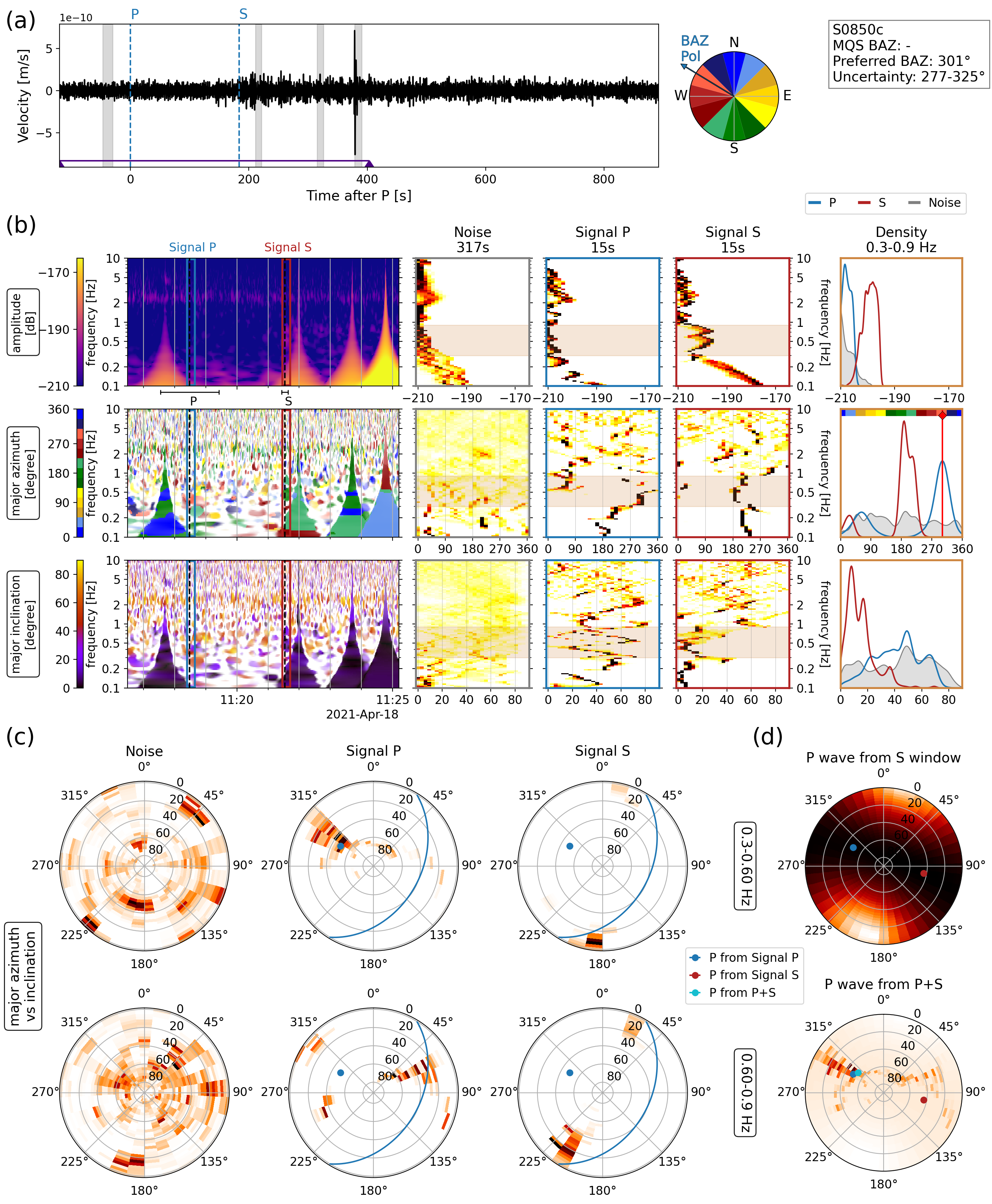}
\caption{Polarization analysis for marsquake S0850c ($M_w$2.6 BB QC event at 31.6$^{\circ}$ distance, April 18, 2021). The plot follows the same structure as Fig. \ref{Fig:Syn_polarization}. There is a strong glitch after the S-wave arrival, though its polarization appears clearly separated from the event signal. The P window shows two dominant azimuths, either at 310\textdegree below 0.6 Hz, or at 60\textdegree above 0.6 Hz. The S window excludes the 60\textdegree direction and instead confirms the range between 277 and 325\textdegree.}
\end{figure}

\begin{figure}[ht]
\centering
\noindent\includegraphics[width=\textwidth]{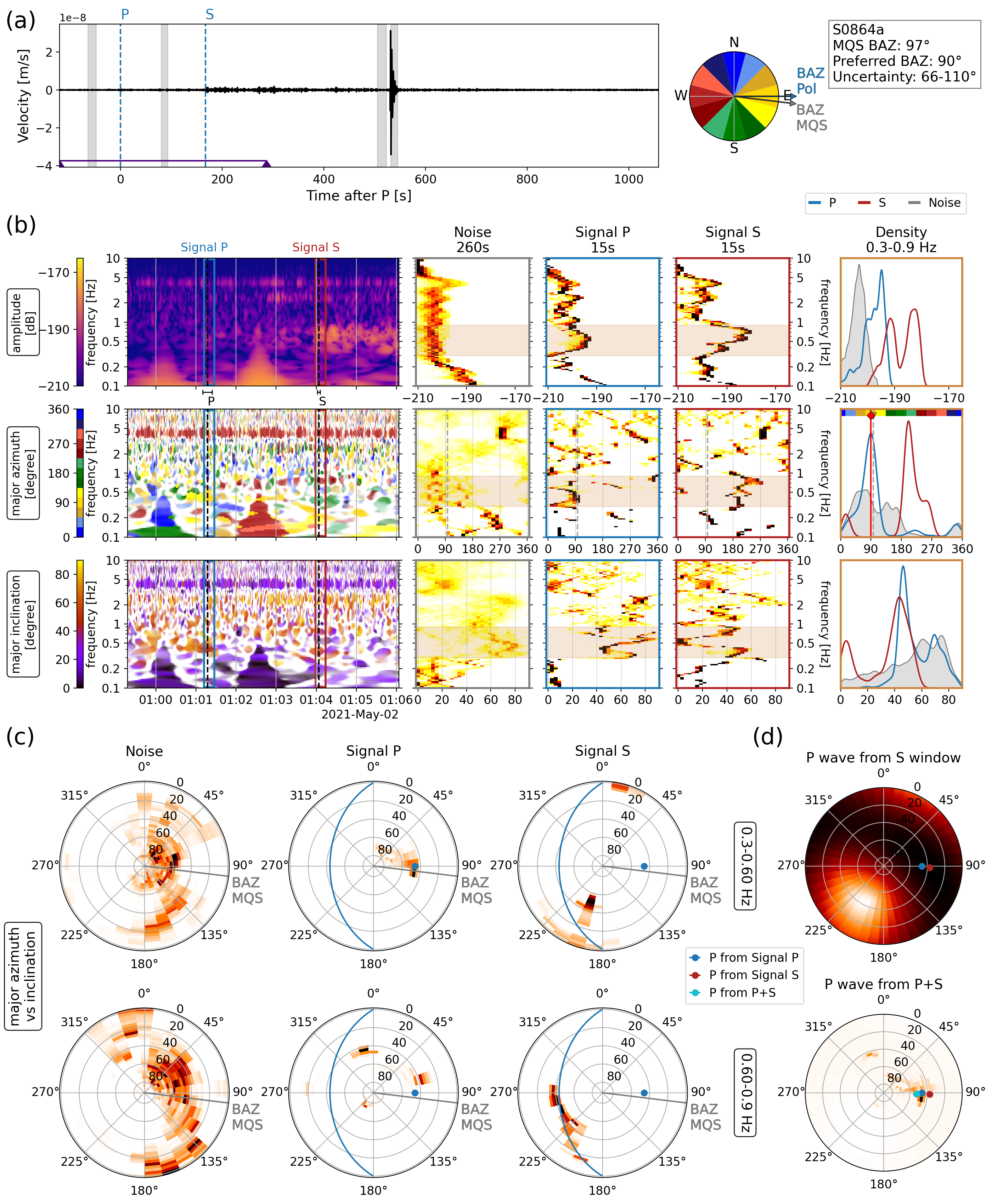}
\caption{Polarization analysis for marsquake S0864a ($M_w$3.1 BB QA event at 28.7$^{\circ}$ distance, May 2, 2021). The plot follows the same structure as Fig. \ref{Fig:Syn_polarization}.  Both P and S arrivals are clear with no glitch contamination. Their amplitude is significantly above the noise and they both show a consistent polarization.}
\end{figure}

\begin{figure}[ht]
\centering
\noindent\includegraphics[width=\textwidth]{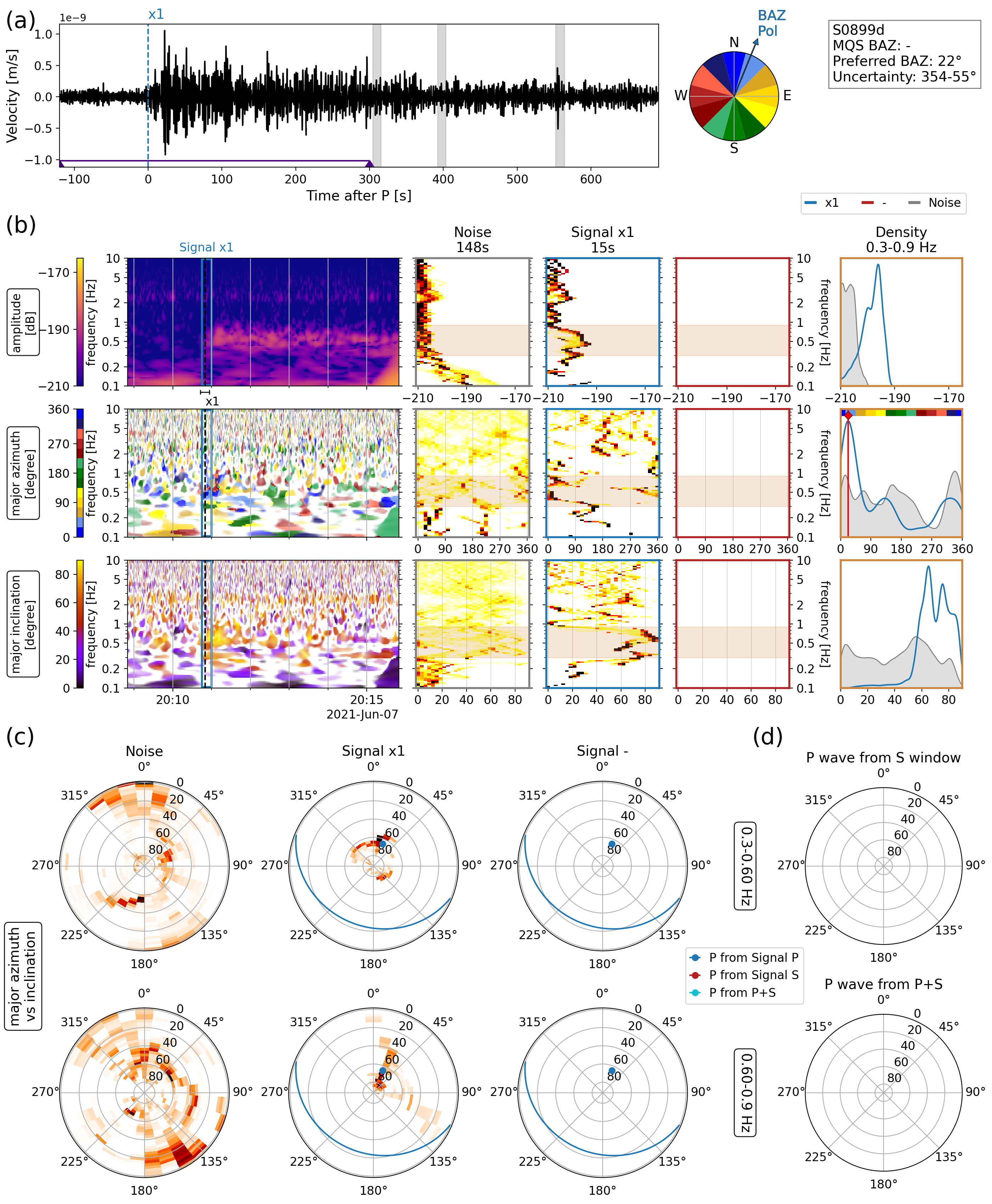}
\caption{Polarization analysis for marsquake S0899d ($M_w$ unknown LF QB event at unknown distance, May 2, 2021). The plot follows the same structure as Fig. \ref{Fig:Syn_polarization}. Since there is no second phase pick, the event has no assigned distance. x1 shows a strongly polarized signal, with a vertical inclination and a back azimuth at 22$^{\circ}$. Though it cannot be corroborated with a second pick, the strong signal amplitude suggests this is the event polarization and not simply noise.}
\end{figure}

\begin{figure}[ht]
\centering
\noindent\includegraphics[width=\textwidth]{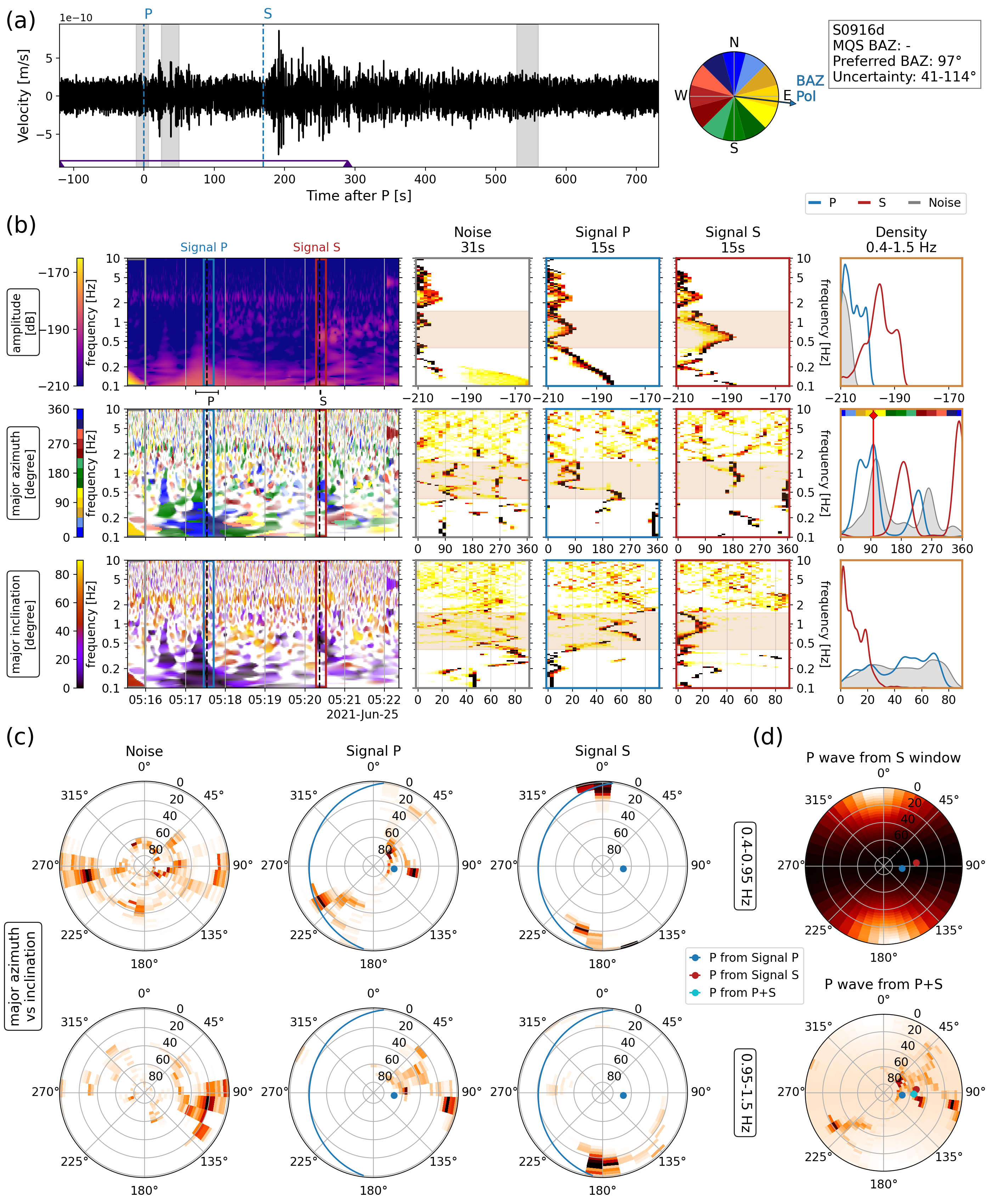}
\caption{Polarization analysis for marsquake S0916d ($M_w$2.9 BB QB event at 29.3$^{\circ}$ distance, June 25, 2021). The plot follows the same structure as Fig. \ref{Fig:Syn_polarization}. The P arrival is contaminated by a strong glitch. Its effect can be minimized by taking a frequency band between 0.4-1.5~Hz. The P-wave azimuth is similar in distribution to the pre-event noies. The S window however shows a clear and strong polarization, with dominant horizontal energy and clear north-south azimuth. Results from both P and S windows agree with each other.}
\end{figure}

\begin{figure}[ht]
\centering
\noindent\includegraphics[width=\textwidth]{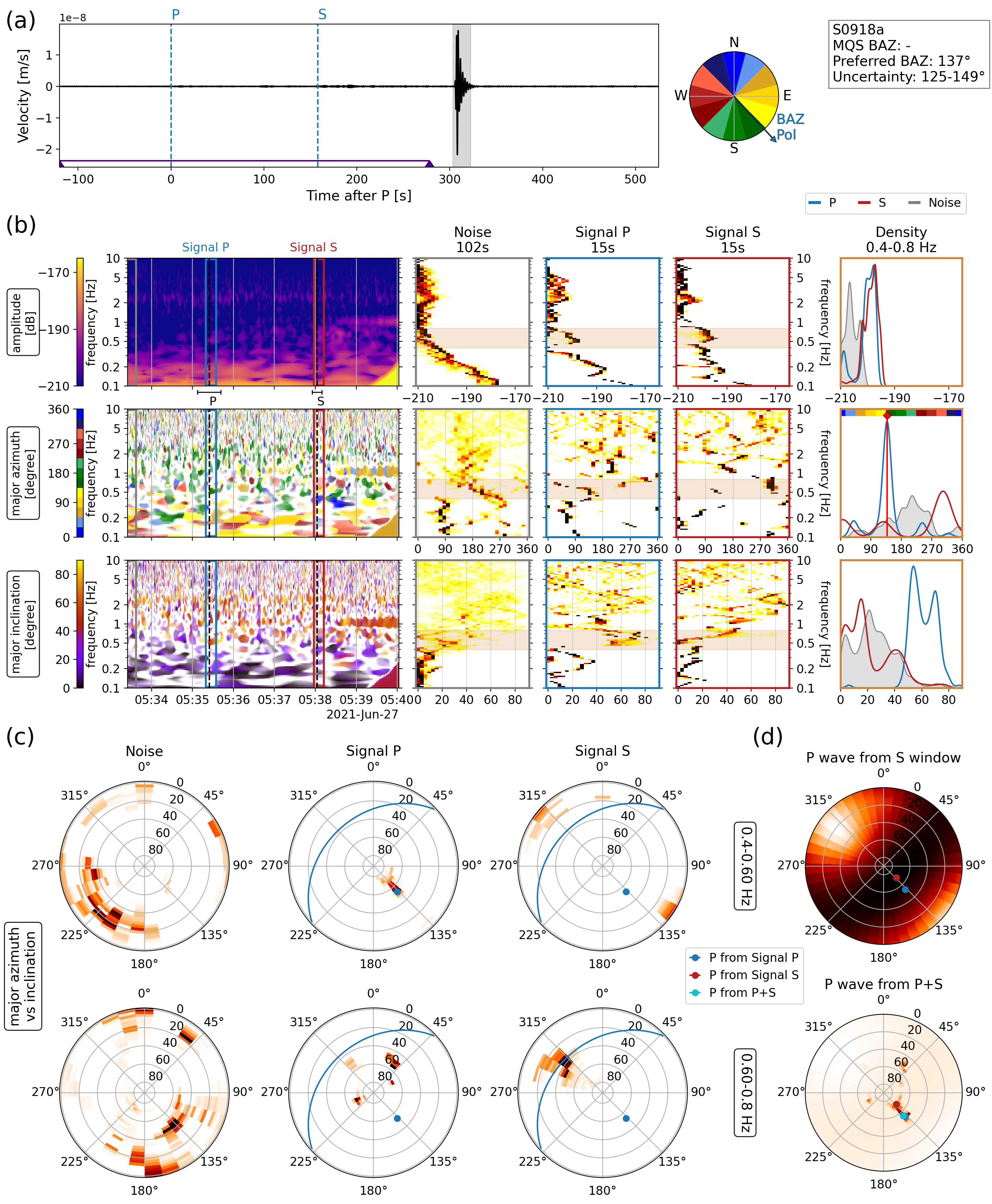}
\caption{Polarization analysis for marsquake S0918a ($M_w$3.0 LF QB event at 27.9$^{\circ}$ distance, June 27, 2021). The plot follows the same structure as Fig. \ref{Fig:Syn_polarization}. Both P and S signals show amplitudes above the noise. Stable P and S polarization is seen between 0.4--0.8~Hz, where P has high inclination and S has low inclination. The P-wave polarization is bimodal, with peaks at 50\textdegree and 135\textdegree, an ambiguity that is not entirely resolved by the S-wave.}
\end{figure}

\begin{figure}[ht]
\centering
\noindent\includegraphics[width=\textwidth]{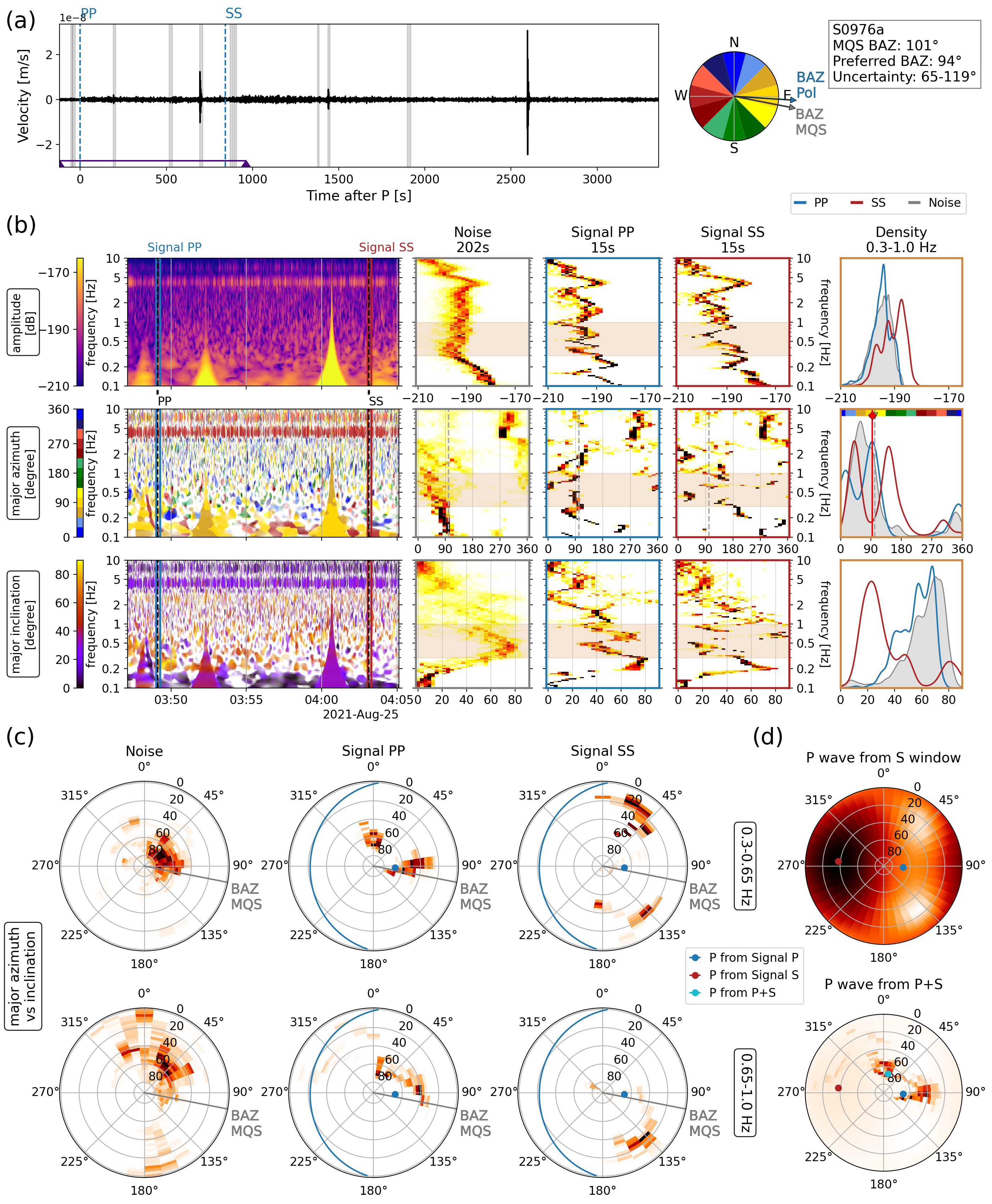}
\caption{Polarization analysis for marsquake S0976a ($M_w$4.2 LF QA event at 146$^{\circ}$ distance, August 25, 2021). The plot follows the same structure as Fig. \ref{Fig:Syn_polarization}. This is the most distant event recorded by InSight so far. The P window polarization shows a consistent back azimuth around 94$^{\circ}$. The S window suggests a back azimuth in opposite direction around 270$^{\circ}$.}
\end{figure}

\begin{figure}[ht]
\centering
\noindent\includegraphics[width=\textwidth]{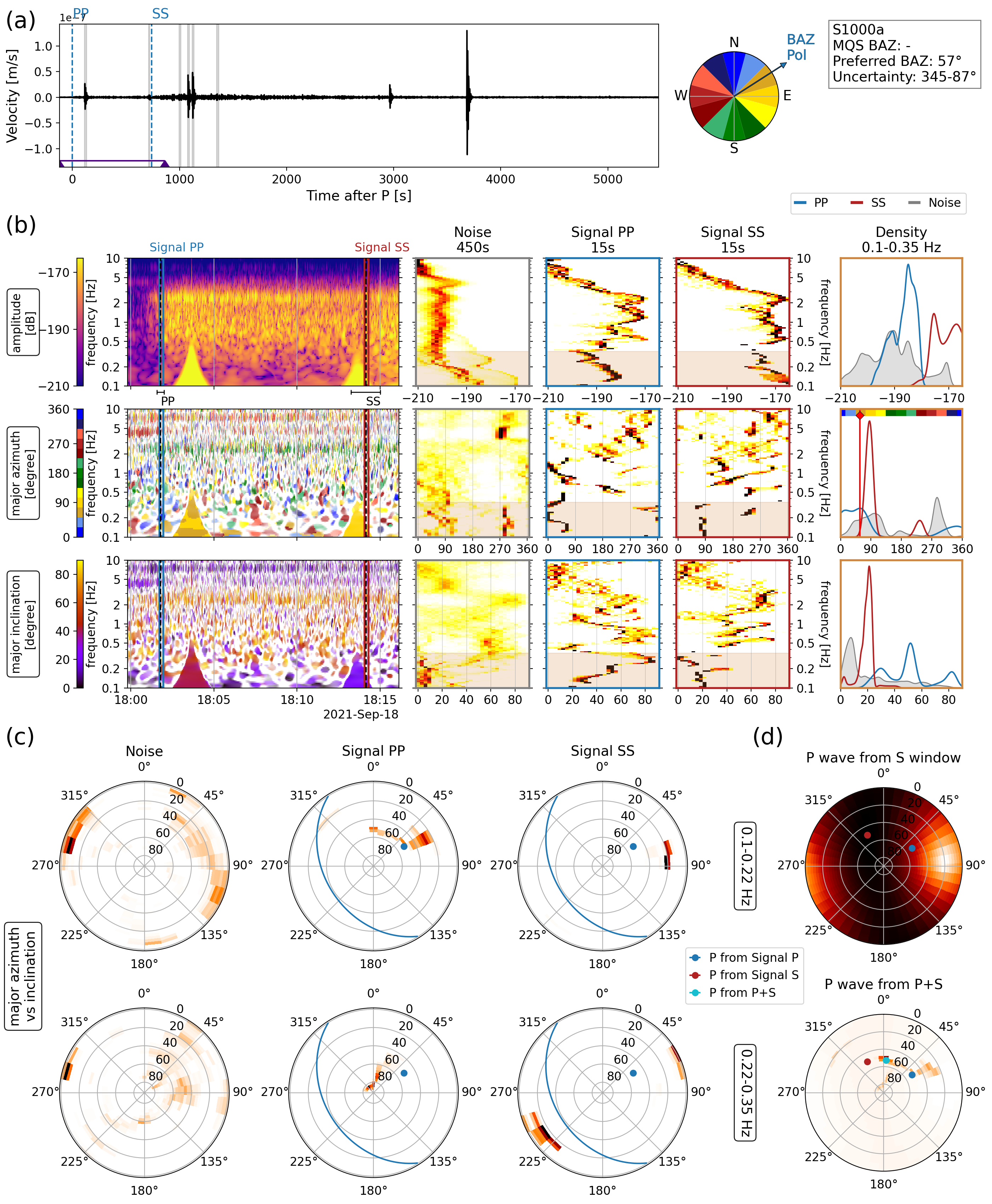}
\caption{Polarization analysis for marsquake S1000a ($M_w$4.1 BB QB event at 116$^{\circ}$ distance, September 18, 2021). The plot follows the same structure as Fig. \ref{Fig:Syn_polarization}. The P window shows a dominant azimuth over a wide frequency band, in northeast direction. Taking a frequency band of 0.1--0.35~Hz puts the back azimuth at 57$^{\circ}$. In this band, the S window is the most horizontal The S window suggests a more north-northwest direction. However, some of that is due to glitch contamination.}
\end{figure}
%

\end{document}